\documentclass[12pt]{article}
\usepackage{setspace}
\usepackage{graphicx}

\usepackage{color}
\input colornam.sty
\begin{document}

\newcommand{\ra}{\rangle }
\newcommand{\la}{\langle }
\newcommand{\ket}[1]{| #1 \rangle }
\newcommand{\bra}[1]{\langle #1 \!| }
\newcommand{ \ave}[2]{ \langle #1 | #2 | #1 \rangle }
\newcommand{\amp }[2]{\langle #1|#2 \rangle }
\newcommand{\weakv}[3]{ \frac{\langle #1|#2| #3\rangle}{\langle #1 | #3 \rangle }}
\newcommand{\beq}{\begin{equation}}
\newcommand{\eeq}{\end{equation}}
\newcommand{\up}[1]{^{(#1)}}
\newcommand{\upn}{^{\mathrm{(N)}}}
\newcommand{\upa}{\uparrow}
\newcommand{\dwa}{\downarrow}

\title{ \bf
\Large Robust Weak Measurements on Finite Samples}

\author{\small Jeff Tollaksen}

\maketitle
\centerline{Center for Quantum Studies }
\centerline{Department of Computational and Data Sciences}
\centerline{Department of Physics}
\centerline{College of Science, George Mason University, Fairfax, VA 22030}

\begin{abstract}

A new weak measurement procedure is introduced for finite samples which yields accurate weak values that are outside the range of eigenvalues and which do not require an exponentially rare ensemble.
This procedure provides a unique 
 advantage in the amplification of small non-random signals by minimizing uncertainties in determining the weak value and by minimizing sample size.
This procedure can also extend the strength of the coupling between the system and measuring device to a new regime.
\end{abstract}


\section{\textcolor{black}{\bf  Introduction}}

Aharonov, Bergmann and Lebowitz
(ABL,~\cite{abl}) considered  measurement situations {\em between} two successive ideal measurements where the transition from a pre-selected state $\ket{\Psi_{\mathrm{in}}}$ to a post-selected state $\ket{\Psi_{\mathrm{fin}}}$ is  generally disturbed by an intermediate precise measurement.  
A subsequent theoretical development arising out of the ABL work was the introduction of the ``Weak Value" (WV) of an observable which was probed by a  new type of quantum measurement called the ``Weak Measurement" (WM) \cite{av} (reviewed in \S \ref{der2v}). 
The motivation behind these measurements was to explore the relationship between $\ket{\Psi_{\mathrm{in}}}$ and $\ket{\Psi_{\mathrm{fin}}}$  by reducing the disturbance on the system during the intermediate time.  This can be modeled  by reducing the interaction strength between the system and the measuring device.  For example, if a WM of $\hat{A}$ is performed at the intermediate time $t$ ($t_{\mathrm{in}}<t<t_{\mathrm{fin}}$) then, in contrast to the ABL situation,  the basic object in the entire interval $t_{\mathrm{in}}\rightarrow t_{\mathrm{fin}}$ for the purpose of calculating {\it other} WVs for other measurements 
is the pair of states $\ket{\Psi_{\mathrm{in}}}$ and $\ket{\Psi_{\mathrm{fin}}}$.   
However, the reduction of disturbance also reduced the information obtained from a single WM on a single  quantum system.  Therefore,  the WV was determined by using a large ensemble (reviewed in \S \ref{wvstat} and \S\ref{wvavgop}).
This was a result of the weakness condition which produced a shift in the pointer of the measuring device (MD) that was much less than the uncertainty. Many separate irreversible recordings of the slight MD shift were then used to amplify the ``weak value signal" above the ``noise" due to the weakened measurement.

This article introduces (\S \ref{rwm}) a new Gedanken experiment coined ``Robust Weak Measurements on Finite Samples" (RWM).  The primary advantage of RSM is a reduction in the uncertainty of the  WV for finite samples and also increases the probability to obtain WVs which are outside the eigenvalue spectrum.
RWM involves an irreversible recording of  the sum of momenta for an ensemble of quantum systems, such that the shift in the sum of momenta  is large compared to it's noise. 
In addition, 
in order to ascertain the maximally allowed information about the WV, 
the relative positions (which commute with the total momenta) are also measured.  
In a physical, ``realistic," WM,  there is always a finite coupling and thus a disturbance caused to the system in addition to unknown fluctuations.  
However, we can use the relative positions to correct for this disturbance and unknown fluctuations and thus WVs can now be determined much more accurately for a finite-sized ensemble.  
We illustrate RWM  by a practical application of WVs to the amplification of weak or unknown signals and show how the interaction strength $\lambda$ can be increased and yet still have a useful regime of WVs.  
RWM also allows for a  reduction in the number of particles necessary to perform an accurate WM.

\label{tpwm}

\section{Weak Measurements}

\label{der2v}

WMs~\cite{av,duck} can be quantified in the quantum measurement theory developed by von Neumann~\cite{vn}.
First we consider an ideal measurement of observable $\hat{A}$ by using an interaction Hamiltonian $H_{\mathrm{int}}$ of the form 
$H_{\mathrm{int}}=-\lambda  \delta(t)\hat{Q}_{\mathrm{md}}\hat{A}$
where $\hat{Q}_{\mathrm{md}}$ is an observable of the MD (e.g. the position of the pointer), $\lambda $ is a coupling constant which determines the strength  of the measurement, and $\delta(t)$ determines the duration of the measurement (setting $\hbar=1$).
For an impulsive  measurement we need the coupling to be strong and the duration short and thus take $\delta (t)$ to be non-zero only for a short time around the moment of interest such that $\int_0^T \delta  (t)dt=1$ (which thereby allows us to ignore the free Hamiltonians of the system and MD).
Using the Heisenberg equations of motion for the momentum $\hat{P}_{\mathrm{md}}$ of MD (conjugate to the position $\hat{Q}_{\mathrm{md}}$), we see that $\hat{P}_{\mathrm{md}}$ changes according to $\frac{d\hat{P}_{\mathrm{md}}}{d t}=\lambda  \delta(t) \hat{A}$.
Integrating this, we see that $P_{\mathrm{md}}(T)-P_{\mathrm{md}}(0)=\lambda \hat{A}$, where $P_{\mathrm{md}}(0)$ characterizes the initial state of MD and $P_{\mathrm{md}}(T)$ characterizes the final.  
To make a more precise determination of $\hat{A}$ requires that the shift in $P_{\mathrm{md}}$, i.e. $\delta P_{\mathrm{md}}=P_{\mathrm{md}}(T)-P_{\mathrm{md}}(0)$, be distinguishable from it's uncertainty, $\Delta P_{\mathrm{md}}$.  This occurs, e.g., if $P_{\mathrm{md}}(0)$ and $P_{\mathrm{md}}(T)$ are more precisely defined and/or if $\lambda$ is sufficiently large.  
However, under these conditions (e.g. if MD approaches a delta function in $P_{\mathrm{md}}$),
 the disturbance or back-reaction on the system is increased due to a larger  $H_{\mathrm{int}}$ which is a result of the larger $\Delta Q_{\mathrm{md}}$ arising from the inverse relationship of $\Delta P_{\mathrm{md}}$ and $\Delta Q_{\mathrm{md}}$ ($\Delta Q_{\mathrm{md}}\geq\frac{1}{\Delta P_{\mathrm{md}}}$).
When $\hat{A}$ is 
measured in this way, then any operator $\hat{O}$ ($[\hat{A},\hat{O}]\neq 0$) is disturbed because $\frac{d}{dt}{\hat{O}}=i\lambda  \delta(t)[\hat{A},\hat{O}]\hat{Q}_{\mathrm{md}}$, and since $\lambda\Delta Q_{\mathrm{md}}$ is not zero, $\hat{O}$ changes in an uncertain way proportional to $\lambda\Delta Q_{\mathrm{md}}$.  

In the Schroedinger picture of measurement, the system and MD state are:
\begin{equation}
|\Phi_{tot}\ra = |\Psi_{\mathrm{in}}\ra |\Phi^{\mathrm{in}}_{\mathrm{md}}\ra\rightarrow e^{ -i  \int H_{\mathrm{int}} dt }|\Psi_{\mathrm{in}}\ra |\Phi^{\mathrm{in}}_{\mathrm{md}}\ra=e^{ i  \lambda\hat{Q}_{\mathrm{md}} \hat{A}}|\Psi_{\mathrm{in}}\ra |\Phi^{\mathrm{in}}_{\mathrm{md}}\ra
\label{measurement2}
\end{equation}
where  the state of the system is $|\Psi_{\mathrm{in}}\ra$ and the MD state, $|\Phi^{\mathrm{in}}_{\mathrm{md}}\ra$, is given by $|\Phi^{\mathrm{in}}_{\mathrm{md}}\ra=\int dQ_{\mathrm{md}} \Phi^{\mathrm{in}}_{\mathrm{md}}(Q_{\mathrm{md}})|Q_{\mathrm{md}}\ra=\int dP \tilde{\Phi}_{in}^{MD}(P_{\mathrm{md}})|P_{\mathrm{md}}\ra$.
A good approximation for realistic experiments is  to consider MD's initial state as a Gaussian (without loss of generality), e.g. 
 $\Phi^{\mathrm{in}}_{\mathrm{md}}(Q_{\mathrm{md}})\equiv\la Q_{\mathrm{md}}|\Phi^{\mathrm{in}}_{\mathrm{md}}\ra = \exp(-{{Q_{\mathrm{md}}^2}\over{4\Delta^2}})$ and $\tilde{\Phi}_{in}^{MD}(P_{\mathrm{md}})\equiv\la P|\Phi^{\mathrm{in}}_{\mathrm{md}}\ra = \exp(-\Delta^2P^2)$ (substituting $\Delta\equiv\Delta
Q_{\mathrm{md}} $,  $\Delta P_{\mathrm{md}}\equiv\frac{1}{\Delta}$, leaving off the normalizations).  Expanding $|\Psi_{\mathrm{in}}\ra$ in eigenstates of $\hat{A}$, i.e. $|\Psi_{\mathrm{in}}\ra =\sum_i|A=a_i\ra\la
 A=a_i|\Psi_{\mathrm{in}}\ra =\sum_i a_i|A=a_i\ra$
then eq. \ref{measurement2}, becomes:
\begin{equation}
  \sum_ia_i\int dQ_{\mathrm{md}} e^{i\lambda \hat{Q}_{\mathrm{md}}a_i}e^{-{{Q_{\mathrm{md}}^2}\over{4\Delta^2}}}|A=a_i\ra|Q_{\mathrm{md}}\ra =\sum_ia_i\int dP e^{{-(\hat{P}-\lambda a_i)^2}\Delta^2}|A=a_i\ra|P\ra 
 \label{measurement3}
\end{equation}
When the uncertainty, $\Delta P_{\mathrm{md}}=\frac{1}{\Delta Q_{\mathrm{md}}}$, in MD is
 much smaller than the shift of MD, $\delta P_{\mathrm{md}}=\lambda a_i$,
 corresponding to the strength of interaction and to different eigenvalues, then the final state of 
 MD is a density matrix representing a series of peaks, each corresponding to a
 different eigenvalue $a_i$, i.e. after tracing over the state of the measured system, the absolute square of eq. \ref{measurement3} yields $Pr(P_{\mathrm{md}})=\sum_i |a_i|^2 e^{-{ 2\Delta^{2}(P_{\mathrm{md}}-\lambda a_i)^{2} }} $.  In other words, MD goes into a state of superposition proportional to the system.  If ${ \Phi^{\mathrm{fin}}_{\mathrm{md}}(P_{\mathrm{md}}-\lambda a_i)}$ is orthogonal to  ${ \Phi^{\mathrm{fin}}_{\mathrm{md}}(P_{\mathrm{md}}-\lambda a_j)}$ when
$i\neq j$ (which occurs if $\delta P_{\mathrm{md}}\gg\Delta P_{\mathrm{md}}$), then this macroscopic superposition collapses into a single peak with probability given by the Born rule $Pr(a_j,t |\Psi_{\mathrm{in}},t_{\mathrm{in}}) = |\bra{a_j} U_{t_{\mathrm{in}}\rightarrow t}\ket{\Psi_{\mathrm{in}}}|^2\equiv|a_j|^2 $, i.e. depending only on the initial state of the measured system.

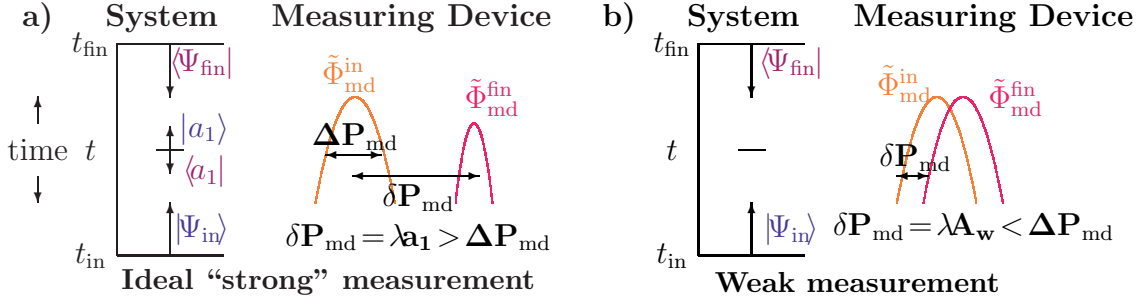
\begin{figure}[h]
\begin{picture}(500,90)(0,0)
\put(40,10){\line(0,1){80}}
\put(40,10){\line(1,0){40}}
\put(55,50){\line(1,0){10}}
\put(60,50){\vector(0,1){9}}
\put(60,50){\vector(0,-1){9}}
\put(40,90){\line(1,0){40}}
\put(60,10){\vector(0,1){20}}
\put(60,90){\vector(0,-1){20}}
\put(72,20){\makebox(0,0){\textcolor{BlueViolet}{$|\!\Psi_{\mathrm{in}}\!\rangle$}}}
\put(72,83){\makebox(0,0){\textcolor{RedViolet}{$\langle\!\Psi_{\mathrm{fin}}\!\!\mid$}}}
\put(73,42){\makebox(0,0){\textcolor{RedViolet}{$\langle\!a_1\!\!\mid$}}}
\put(73,58){\makebox(0,0){\textcolor{BlueViolet}{$|a_1\rangle$}}}
\put(10,50){\makebox(0,0){time}}
\put(30,10){\makebox(0,0){$t_{\mathrm{in}}$}}
\put(30,90){\makebox(0,0){$t_{\mathrm{fin}}$}}
\put(30,50){\makebox(0,0){$t$}}
\put(10,40){\vector(0,-1){10}}
\put(10,60){\vector(0,1){10}}
\put(10,100){\makebox(0,0){\bf a)}}
\put(120,100){\makebox(0,0){\bf System \,\,\,\,\,\,\,Measuring Device}}
\put(120,0){\makebox(0,0){\bf \small Ideal ``strong" measurement}}

\put(128,79){\makebox(0,0){{\textcolor{Orange}{\bf $\tilde{\Phi}^{\mathrm{in}}_{\mathrm{md}}$}}}}

\color{Orange} 

\bezier{500}(115,30)(130,110)(145,30)

\color{WildStrawberry}
\put(182,70){\makebox(0,0){\bf $\tilde{\Phi}^{\mathrm{fin}}_{\mathrm{md}}$}}

\bezier{500}(168,30)(175,90)(182,30)

\color{black}
\put(154,32){\makebox(0,0){$\bf \delta P_{\mathrm{md}}$}}
\put(130,48){\vector(1,0){10}}
\put(130,48){\vector(-1,0){11}}
\put(130,55){\makebox(0,0){$\bf \Delta P_{\mathrm{md}}$}}

\put(154,40){\vector(1,0){23}}
\put(154,40){\vector(-1,0){25}}
\put(154,17){\makebox(0,0){$\bf \delta P_{\mathrm{md}}\!=\!\lambda \!a_1\!>\!\Delta P_{\mathrm{md}}$}}


\put(338,75){\makebox(0,0){{\textcolor{Orange}{\bf $\tilde{\Phi}^{\mathrm{in}}_{\mathrm{md}}$}}}}

\color{Orange} 

\bezier{500}(335,30)(350,110)(365,30)

\color{WildStrawberry}
\put(380,70){\makebox(0,0){\bf $\tilde{\Phi}^{\mathrm{fin}}_{\mathrm{md}}$}}

\bezier{500}(345,30)(360,110)(375,30)


\color{black}
\put(342,47){\makebox(0,0){$\bf \delta P_{\mathrm{md}}$}}
\put(341,40){\vector(1,0){6}}
\put(341,40){\vector(-1,0){6}}
\put(364,21){\makebox(0,0){$\bf \delta P_{\mathrm{md}}\!=\!\lambda \!A_w\!<\!\Delta P_{\mathrm{md}}$}}

\put(320,0){\makebox(0,0){\bf \small Weak measurement}}
\put(230,100){\makebox(0,0){\bf b)}}
\put(340,100){\makebox(0,0){\bf System \,\,\,\,\,\,\,Measuring Device}}

\put(260,10){\line(0,1){80}}
\put(260,10){\line(1,0){40}}
\put(275,50){\line(1,0){10}}
\put(260,90){\line(1,0){40}}
\put(280,10){\vector(0,1){20}}
\put(280,90){\vector(0,-1){20}}
\put(295,20){\makebox(0,0){\textcolor{BlueViolet}{$|\!\Psi_{\mathrm{in}}\!\rangle$}}}
\put(295,83){\makebox(0,0){\textcolor{RedViolet}{$\langle\!\Psi_{\mathrm{fin}}\!\!\mid$}}}
\put(250,10){\makebox(0,0){$t_{\mathrm{in}}$}}
\put(250,90){\makebox(0,0){$t_{\mathrm{fin}}$}}
\put(250,50){\makebox(0,0){$t$}}

\end{picture}

\caption[Two vector formulation with strong and weak measurements]{\small  a) 
with an ideal or ``strong" measurement at $t$ (characterized e.g. by $\delta P_{\mathrm{md}}\!=\!\lambda \!a_1\!\gg\!\Delta P_{\mathrm{md}}$), then ABL gives the probability to obtain a collapse onto eigenstate $a_1$ by propagating \textcolor{RedViolet}{$\langle\!\Psi_{\mathrm{fin}}\!\!\mid$} backwards in time from $t_{\mathrm{fin}}$
to $t$ and \textcolor{BlueViolet}{$|\!\Psi_{\mathrm{in}}\!\rangle$} forwards in time from $t_{\mathrm{in}}$ to $t$; in addition, the collapse caused by the ideal measurement at $t$ creates a new boundary condition 
\textcolor{BlueViolet}{$|a_1\rangle$}
$\!\!$\textcolor{RedViolet}{$\langle\!a_1\!\!\mid$} at time $t$, $t_{\mathrm{in}}<t<t_{\mathrm{fin}}$; b) 
if a WM is performed at $t$ (characterized e.g. by $\delta P_{\mathrm{md}}\!=\!\lambda \!A_w\!\ll\!\Delta P_{\mathrm{md}}$), then the outcome of the WM, the WV, can be calculated by propagating the
state \textcolor{RedViolet}{$\langle\!\Psi_{\mathrm{fin}}\!\!\mid$} backwards in time from $t_{\mathrm{fin}}$
to $t$ and the state \textcolor{BlueViolet}{$|\!\Psi_{\mathrm{in}}\!\rangle$} 
forwards in time from 
$t_{\mathrm{in}}$ to $t$; the WM does not cause a collapse and thus no new boundary condition is created at time $t$.}
\label{seqm1}
\end{figure}

However, suppose at a later time $t_{\mathrm{fin}}$, we measure another non-degenerate operator which yields, as one of its outcomes, the state $\ket{\Psi_{\mathrm{fin}}}$. The conditional probability to obtain $a_j$, given both boundary conditions, the pre-selected $\ket{\Psi_{\mathrm{in}}}$ and  post-selected $\bra{\Psi_{\mathrm{fin}}}$, is given by ABL~\cite{abl}:
\begin{eqnarray}
Pr(a_j,t|\Psi_{\mathrm{in}},t_{\mathrm{in}}; \Psi_{\mathrm{fin}},t_{\mathrm{fin}}) & = & \frac{ Pr(\Psi_{\mathrm{fin}},t_{\mathrm{fin}} |a_j,t) Pr(a_j,t |\Psi_{\mathrm{in}},t_{\mathrm{in}}) }{\sum_{a'} Pr(\Psi_{\mathrm{fin}},t_{\mathrm{fin}}|a',t;\Psi_{\mathrm{in}},t)}\nonumber \\
& = & \frac{ |\bra{\Psi_{\mathrm{fin}}} U_{t\rightarrow t_{\mathrm{fin}}}\ket{a_j}\bra{a_j} U_{t_{\mathrm{in}}\rightarrow t}\ket{\Psi_{\mathrm{in}}}|^2 }{\sum_{a'} |\bra{\Psi_{\mathrm{fin}}} U_{t\rightarrow t_{\mathrm{fin}}}\ket{a'}\bra{a'} U_{t_{\mathrm{in}}\rightarrow t}\ket{\Psi_{\mathrm{in}}}|^2} \!
\label{ablnts}
\end{eqnarray}
To illustrate the time-symmetry in ABL, we apply $U_{t\rightarrow t_{\mathrm{fin}}}$ on $\bra{\Psi_{\mathrm{fin}}}$, i.e. $\bra{\Psi_{\mathrm{fin}}}U_{t\rightarrow t_{\mathrm{fin}}}=\langle U_{t\rightarrow t_{\mathrm{fin}}}^\dag\Psi_{\mathrm{fin}}|$ and use $U_{t\rightarrow
      t_{\mathrm{fin}}}^\dag={\left\{e^{-iH(t_{\mathrm{fin}}-t)}\right\}}^\dag=e^{iH(t_{\mathrm{fin}}-t)}=e^{-iH(t-t_{\mathrm{fin}})}=U_{t_{\mathrm{fin}}\rightarrow t}$.   The time-symmetric re-formulation of the numerator of eq. \ref{ablnts}, $\langle U_{t_{\mathrm{fin}}\rightarrow t}\Psi_{\mathrm{fin}}|a_j\rangle\langle a_j|U_{t_{\mathrm{in}}\rightarrow t}|\Psi_{\mathrm{in}}\rangle$,  can now be interpreted to mean that the time displacement operator $U_{t_{\mathrm{fin}}\rightarrow t}$ sends $\bra{\Psi_{\mathrm{fin}}}$ back in time to $t$ as depicted in fig. \ref{seqm1}.a.  The Born rule, of course, is recovered by summing over all possible final states $\{\Psi_{\mathrm{fin}}\}_n$:
\beq
Pr(a_j,t|\Psi_{\mathrm{in}};t_{\mathrm{in}}) = \sum_{n} Pr(a_j,t|\Psi_{\mathrm{in}},t_{\mathrm{in}};\{\Psi_{\mathrm{fin}}\}_n,t_{\mathrm{fin}})\, Pr(\{\Psi_{\mathrm{fin}}\}_n,t_{\mathrm{fin}}|a_j,t;\Psi_{\mathrm{in}},t) \, .
\eeq

We can weaken the interaction $H_{\mathrm{int}}\!=\!-\lambda  \delta(t)\hat{Q}_{\mathrm{md}}\hat{A}$
 by minimizing $\lambda \Delta Q_{\mathrm{md}}$.  For simplicity, we consider $\lambda\ll 1$ (assuming without lack of generality that the state of the MD is a Gaussian with spreads $\Delta P_{\mathrm{md}}\!=\!\Delta Q_{\mathrm{md}}\!=\!1$).   We may then set $e^{ -i \lambda  \hat{Q}_{\mathrm{md}} \hat{A}
 }\!\approx\! 1-i\lambda \hat{Q}_{\mathrm{md}} \hat{A}$ and use a theorem~\cite{avg}\footnote{where $\la\hat{A}\ra = \ave{\Psi}{\hat{A}}$, $|\Psi\rangle$ is any vector in Hilbert space, $\Delta A^2 = \ave{\Psi}{(\hat{A} - \la\hat{A}\ra)^2}$, and $\ket{\Psi_\perp}$ is a state such that $\amp {\Psi}{\Psi_\perp} = 0$. Note that $\la\hat{A}\ra$ is not defined here in a statistical sense: it is a mathematical property of an individual system $|\Psi\rangle$.}:
\beq\label{identity}
\hat{A} |\Psi \rangle = \la\hat{A}\ra \ket{\Psi}  + \Delta A \ket{\Psi_\perp}\, ,
\label{thm1}
\eeq
 to show that before the post-selection, the system state is:
   \begin{equation}
      e^{ i  \lambda\hat{Q}_{\mathrm{md}} \hat{A}}|\Psi_{\mathrm{in}}\ra\!=\!
(1\!-\!i\lambda \hat{Q}_{\mathrm{md}}\hat{A})|\Psi_{\mathrm{in}}\rangle\!=\!(1\!-i\!\lambda \hat{Q}_{\mathrm{md}}\langle\hat{A}\rangle)|\Psi_{\mathrm{in}}\rangle\!-i\!\lambda \hat{Q}_{\mathrm{md}}\Delta\hat{A}|\Psi_{\mathrm{in}\perp}\rangle
   \end{equation}

Using the norm of this state ${\parallel (1-i\lambda \hat{Q}_{\mathrm{md}}\hat{A})|\Psi_{\mathrm{in}}\rangle
      \parallel}^2=1+{\lambda ^2\hat{Q}_{\mathrm{md}}^2}\langle \hat{A}^2\rangle$,
 the probability to leave $|\Psi_{\mathrm{in}}\rangle$ un-changed after the measurement is:
   \begin{equation}
      \frac{1+{\lambda ^2\hat{Q}_{\mathrm{md}}^2}{\langle\hat{A}\rangle}^2}
   {1+{\lambda ^2\hat{Q}_{\mathrm{md}}^2}\langle
   \hat{A}^2\rangle}\longrightarrow 1\,\,\,\,\,\,(\lambda \rightarrow 0)
   \end{equation}
while the probability to disturb the state (i.e. to obtain $|\Psi_{in\perp}\rangle$) is:
   \begin{equation}\label{14.11}
      \frac{{\lambda ^2\hat{Q}_{\mathrm{md}}^2}{\Delta\hat{A}}^2}
   {1+{\lambda ^2\hat{Q}_{\mathrm{md}}^2}\langle
   \hat{A}^2\rangle}\longrightarrow 0\,\,\,\,\,\,(\lambda \rightarrow 0)
\label{collprob}
   \end{equation}
The probability for a collapse  decreases as $\frac{1}{\lambda ^2}$.  Thus, 
for a sufficiently weak interaction (e.g. $\lambda\ll 1$), the probability for a collapse can be made arbitrarily small.  In addition, the measurement becomes less precise because 
the shift in MD is much smaller than its uncertainty (i.e. $\delta P_{\mathrm{md}}=\lambda a_i\ll\Delta P_{\mathrm{md}}$).
The final state of MD is now a superposition of many substantially overlapping Gaussians with the same distribution as before $Pr(P_{\mathrm{md}})=\sum_i |\la A=a_i|\Psi_{\mathrm{in}}\ra|^2 \exp\left\{{-\frac{(P_{\mathrm{md}}-\lambda a_i)^{2}} {2\Delta P_{\mathrm{md}}^{2}}}\right\} $.  However, this ends up being a single Gaussian 
 $\tilde{\Phi}^{\mathrm{fin}}_{\mathrm{md}}(P_{\mathrm{md}})\approx\langle P_{\mathrm{md}}|e^{-i\lambda \hat{Q}_{\mathrm{md}}\la\hat{A}\ra}|\Phi^{\mathrm{in}}_{\mathrm{md}}\rangle\approx\exp\left\{-{{(P_{\mathrm{md}}-\lambda\la\hat
 A\ra)^2}\over{\Delta P_{\mathrm{md}}^2}}\right\}$ centered on $\lambda\la\hat{A}\ra$.

If we perform this measurement on a single particle, then, of course, we will not be able to distinguish between two states which are not orthogonal, e.g. $|\Psi_1\ra|\Phi^{\mathrm{in}}_{\mathrm{md}}\ra$ and $|\Psi_2\ra|\Phi^{\mathrm{in}}_{\mathrm{md}}\ra$.  Such an ability would violate unitarity because these states could time evolve into orthogonal states $|\Psi_1\ra|\Phi^{\mathrm{in}}_{\mathrm{md}}\ra \longrightarrow |\Psi_1\ra|\Phi^{\mathrm{in}}_{\mathrm{md}}(1)\ra$ and 
$|\Psi_2\ra|\Phi^{\mathrm{in}}_{\mathrm{md}}\ra \longrightarrow |\Psi_2\ra|\Phi^{\mathrm{in}}_{\mathrm{md}}(2)\ra$, with $|\Psi_1\ra|\Phi^{\mathrm{in}}_{\mathrm{md}}(1)\ra$ orthogonal to $|\Psi_2\ra|\Phi^{\mathrm{in}}_{\mathrm{md}}(2)\ra$.  From a WM perspective, the reason that this does not happen is that measurement of these two non-orthogonal states causes a smaller shift in MD than it's uncertainty and therefore we might conclude that the shift $\delta P_{\mathrm{md}}$ of MD is  a measurement error because $ \tilde{\Phi}_{\mathrm{fin}}^{\mathrm{MD}}(P_{\mathrm{md}})=\langle
      P_{\mathrm{md}}-\lambda \la\hat{A}\ra|\Phi^{\mathrm{in}}_{\mathrm{md}}\rangle\approx\langle P_{\mathrm{md}}|\Phi^{\mathrm{in}}_{\mathrm{md}}\rangle$ for $\lambda \ll 1$.  
Nevertheless, if a large ($N\geq\frac{N'}{\lambda }$) ensemble of  particles is used, then the shift of all the MDs ($\delta P^{tot}_{\mathrm{md}}\approx\lambda \la\hat{A}\ra\frac{N'}{\lambda}=N'\la\hat{A}\ra$) can accumulate to a distinguishable level while  the collapse probability   still goes to zero.
That is, for a large ensemble of particles which are all either $|\Psi_2\ra$ or
   $|\Psi_1\ra$, this measurement can distinguish between them even if $|\Psi_2\ra$ and
   $|\Psi_1\ra$ are not orthogonal because the scalar product $\langle\Psi_1\upn|\Psi_2\upn\rangle=\cos^n\theta\longrightarrow 0$.
Traditionally, it was believed that if a measurement interaction is limited so there is no disturbance on the system, then no information is gained. 
However, we have shown that when considered as a limiting process, the disturbance goes to zero more quickly than the shift in MD and thus with a large enough ensemble, information can be obtained even though not even a single particle was disturbed.

Now that we have a new measurement which does not cause a collapse, we ask whether this type of measurement might reveal new values.  We shall see that with a WM (which involves adding a post-selection to this weakened von Neumann measurement),  the MD registers a new value, the WV.
As an indication of this, we insert a complete set of states $\{ \ket{\Psi_{\mathrm{fin}}}_j \}$ into 
$\la\hat{A}\ra$:
\beq
 \la\hat{A}\ra =  \bra{\Psi_{\mathrm{in}}} { \left[\sum_j  \ket{\Psi_{\mathrm{fin}}}_j\bra{\Psi_{\mathrm{fin}}}_j\right]\hat{A}} \ket{\Psi_{\mathrm{in}}}
= \sum_j |\langle \Psi _{\mathrm{fin}} \!\mid_j \!\Psi _{\mathrm{in}}
\rangle|^2\ 
{ {\langle \Psi _{\mathrm{fin}}\! \mid_j \hat{A} \mid \!\Psi _{\mathrm{in}}
\rangle} \over {\langle \Psi _{\mathrm{fin}} \!\mid_j \!\Psi _{\mathrm{in}}
\rangle}}
\label{expweak}
\eeq
If we interpret the states $ \ket{\Psi_{\mathrm{fin}}}_j $ as the possible outcomes of a final ideal measurement on the system (i.e. a post-selection)
and we perform a WM (e.g. with $\lambda\Delta Q_{\mathrm{md}}\rightarrow 0$) during the intermediate time $t$, $t_{\mathrm{in}}<t<t_{\mathrm{fin}}$, then the coefficients $|\langle \Psi_{\mathrm{fin}}\! |_j \Psi_{\mathrm{in}}
\rangle |^2$ give  the probabilities $Pr(j)$ 
for obtaining a pre-selection of $\bra{\Psi_{\mathrm{in}}}$ and a post-selection of  $ \ket{\Psi_{\mathrm{fin}}}_j $ (since the intermediate WM does not disturb these states) 
and the  quantity
$A_{\mathrm{w}}(j) \equiv { {\langle \Psi _{\mathrm{fin}} \!\mid_j \hat{A} \mid \!\Psi _{\mathrm{in}}
\rangle} \over {\langle \Psi _{\mathrm{fin}} \!\mid_j \!\Psi _{\mathrm{in}}
\rangle}}$
 is the WV of $\hat{A}$ given a particular final post-selection $\langle \Psi _{\mathrm{fin}} \!\mid_j$. 
Thus, from 
$\la\hat{A}\ra = \sum_j Pr(j)\,  A_{\mathrm{w}}(j)$,
one can think of  $\la\hat{A}\ra$ for the whole ensemble as being built out of pre- and post-selected
states in which  the WV is multiplied by a probability for post-selection.

To see how the WV arises from a weakened measurement with post-selection more precisely,  we 
consider the final state of MD in the position representation: 

\begin{eqnarray}
{\Phi}_{\mathrm{fin}}^{\mathrm{MD}}(Q_{\mathrm{md}}) &= & 
\la Q_{\mathrm{md}}|\la\Psi_{\mathrm{fin}}|\Phi_{tot}\ra = \la\Psi_{\mathrm{fin}}|e^{ i  \lambda\hat{Q}_{\mathrm{md}} \hat{A}}|\Psi_{\mathrm{in}}\ra e^{-{{Q_{\mathrm{md}}}^{2} \over {4\Delta^{2}}}} \nonumber \\
&=& \sum_{n=0}^{\infty} \frac{(-i\lambda\hat{Q}_{\mathrm{md}})^n}{n!} \langle\Psi_{\mathrm{fin}}\!| \hat{A}^n
|\Psi_{\mathrm{in}} \rangle e^{-{{Q_{\mathrm{md}}}^{2} \over {2\Delta^{2}}}}  
=\langle\Psi_{\mathrm{fin}}\!| \Psi_{\mathrm{in}}\ra\sum_{n=0}^{\infty} \frac{(-i\lambda\hat{Q}_{\mathrm{md}})^n}{n!} (A^n)_{\mathrm{w}}
e^{-{{Q_{\mathrm{md}}}^{2} \over {2\Delta^{2}}}} \nonumber \\
&=& \langle\Psi_{\mathrm{fin}}\!|
\Psi_{\mathrm{in}} \rangle \lbrace 1+i\lambda \hat{Q}_{\mathrm{md}} A_{\mathrm{w}} + \sum_{n=2}^{\infty}
{(i\lambda \hat{Q}_{\mathrm{md}})^{n} \over n!} A^{n}_{\mathrm{w}} \rbrace e^{-{{Q_{\mathrm{md}}}^{2} \over {2\Delta^{2}}}} \nonumber\\
&=& \langle\Psi_{\mathrm{fin}}\!| \Psi_{\mathrm{in}} \rangle \lbrace 
1+i\lambda \hat{Q}_{\mathrm{md}} A_{\mathrm{w}} +  \sum_{n=2}^{\infty} -
{(i\lambda \hat{Q}_{\mathrm{md}})^{n} \over n!} (A_{\mathrm{w}})^{n} \nonumber\\
&\,\,\,\,\,\,\,\,& - \sum_{n=2}^{\infty}
{(i\lambda \hat{Q}_{\mathrm{md}})^{n} \over n!} (A_{\mathrm{w}})^{n} +\sum_{n=2}^{\infty}
{(i\lambda \hat{Q}_{\mathrm{md}})^{n} \over n!} (A^{n})_{\mathrm{w}} \rbrace e^{-{{Q_{\mathrm{md}}}^{2} \over {2\Delta^{2}}}} \nonumber \\
&=& \langle\Psi_{\mathrm{fin}}\!| \Psi_{\mathrm{in}}  \rangle \lbrace e^{-i
\lambda \hat{Q}_{\mathrm{md}}A_{\mathrm{w}}}  +  \sum_{n=2}^{\infty}
{(i\lambda \hat{Q}_{\mathrm{md}})^{n} \over n!} [A^{n}_{\mathrm{w}} - (A_{\mathrm{w}})^{n}] \rbrace e^{-{{Q_{\mathrm{md}}}^{2} \over {2\Delta^{2}}}}
\label{mstate2}
\end{eqnarray}
The second term in the last part of eq. \ref{mstate2}) can be neglected in 2 general regimes: 
\begin{enumerate}
\item minimizing $\lambda\Delta Q_{\mathrm{md}}$ by either using a small $\lambda$ (setting  $\Delta P_{\mathrm{md}}=\Delta Q_{\mathrm{md}}=1$), or by  minimizing the spreads in MD, e.g. so that  $\hat{P}_{\mathrm{md}}$ is measured to a finite
 precision
 $\Delta P_{\mathrm{md}}$, (which limits the disturbance  by a
 finite amount $\Delta Q_{\mathrm{md}}\geq  1/\Delta P_{\mathrm{md}}$), or
\item  minimizing $[A^{n}_{\mathrm{w}} -
(A_{\mathrm{w}})^{n}]/n!$   even if $\lambda\Delta Q_{\mathrm{md}}$ is not small.   
\end{enumerate}
By way of example, the first moment in the Taylor's expansion (from the second term in the last part of eq. \ref{mstate2}) can be neglected if $(\lambda\Delta Q_{\mathrm{md}})^2\Delta A_w\ll 1$ where $\Delta A_w\equiv |(A^2)_w-(A_w)^2|^\frac{1}{2}$~\cite{av}.
When eq. \ref{mstate2} is transformed back to the $P_{\mathrm{md}}$
representation, then 
the final state of MD after WM and post-selection is (up to normalization): 
\begin{eqnarray}
\label{wv0}
 \tilde{\Phi}_{\mathrm{fin}}^{\mathrm{MD}}(P_{\mathrm{md}}) & \approx &  \langle\Psi_{\mathrm{fin}}\ket{\Psi_{\mathrm{in}}}\bra{P_{\mathrm{md}}}e^{ -i \lambda  \hat{Q}_{\mathrm{md}}
A_{\mathrm{w}}
 }\ket{\Phi^{\mathrm{in}}_{\mathrm{md}}}  \approx\exp\left\{{-{{\Delta^2(P_{\mathrm{md}}-\lambda \,
 A_{\mathrm{w}})^2}
}}\right\}\\
where \,\,\,A_w&=&\weakv {\Psi_\mathrm{fin}}{\hat{A}}{\Psi_\mathrm{in} }
\label{wv1}
 \label{post_selected}
\end{eqnarray}
The final
 state
 of MD is almost un entangled with the
 system and has the same initial shape but shifted by a very surprising amount, the WV, $A_{\mathrm{w}}$ (the factor $\langle\Psi_{\mathrm{fin}}\ket{\Psi_{\mathrm{in}}}$ arises as a result of the exclusion of other post-selections).  Since the value of $\hat{A}$ is given by $P_{\mathrm{md}}(T)-P_{\mathrm{md}}(0)$, we may conclude $\hat{A}\approx A_w$.  
We have used such limited
 disturbance measurements to explore many 
 paradoxes (see, e.g. \cite{at2}).  There have also been a number  of experiments to test the predictions made by the WM and their results are in very good agreement with
 theoretical predictions \cite{RSH,Ahnert,Pryde,Wiseman,Parks}.  

The new developments presented in this article are motivated by an  application of WMs and WVs to
quantum metrology 
which can provide a unique 
 advantage over the usual (pre-selected-only) approach in the amplification of small non-random signals.
In the pre-selected-only approach, the outcome of quantum measurements are restricted to the eigenvalue spectrum range. 
The advantage of WVs is that they can be far outside this range due to the overlap $\langle\Psi_{\mathrm{fin}}\ket{\Psi_{\mathrm{in}}}$ in the denominator of $A_{\mathrm{w}}$, eq. \ref{wv1}.   
These WVs are termed ``eccentric weak values" (EWV).  
Now, if the pre- and post-selection are known to high precision, then in the idealized weak limit (e.g. $\lambda\Delta Q_{\mathrm{md}} \rightarrow 0$), the WV 
 can  be calculated precisely 
 and one might conclude that no new information is obtained if a WM is actually performed since MD will simply register a shift by $\lambda A_{\mathrm{w}}$.
However, if the coupling $\lambda$ between system and MD is unknown and contains additional small errors which are not random, then actually performing a WM which yields an EWV can provide new information, e.g. by allowing us to distinguish  the shift from the non-random force (incorporated into $\lambda$) from the large EWV shift due to the WM interaction.  

The intention of this article is to address several theoretical issues concerning such an application. 
In particular, we note that the WV approximations presented above become more and more precise in the idealized weak limit of $\lambda\Delta Q_{\mathrm{md}} \rightarrow 0$ and $N\rightarrow\infty$ in which there is no disturbance or back-reaction on the system.  Neither of these limits are realistic in practical amplifications because first of all, we must have a finite $N$, and second of all with a finite $N$, we must increase $\lambda$ to discern the WV from the noise, and therefore we can no longer ignore the two uncertainties in determining the WV which arise due to:
\begin{enumerate}
\item the inability of MD to measure definite WVs due to the MD's uncertainty $\Delta P_{\mathrm{md}}$  
\item the back-reaction on the system due to $\Delta Q_{\mathrm{md}}$ creates an uncertainty in the WV of the system itself
\end{enumerate} 
 In this article, we demonstrate a new approach to WMs, coined ``Robust Weak Measurements on Finite Samples" (RWM) which decrease these uncertainties and allow us to increase the coupling 
(e.g. $\lambda\sim 1$, not $\lambda \ll 1$ as used in eq. \ref{wv0}) for {\it finite} $N$ while maintaining the accuracy of the WV.  In order to accomplish this, we will have to consider the structure of MD in $Q_{\mathrm{md}}$ (in addition to $P_{\mathrm{md}}$ used above).   
Intuitively, we can see 2 inverse roles for MD observable $\hat{Q}_{\mathrm{md}}$ and system observable $\hat{A}$.  We have already reviewed how $\hat{Q}_{\mathrm{md}}$ generates translations in $\hat{P}_{\mathrm{md}}$ proportional to $A_w$.  However, the roles for $\hat{Q}_{\mathrm{md}}$ and $\hat{A}$ are reversed when one asks the reverse question: ``What is 
the back reaction on the {\it system} (i.e. not on the {\it MD}) due to the measurement interaction $H_{\mathrm{int}}$?"  For this question,  $\hat{A}$ is now the generator (not $\hat{Q}_{\mathrm{md}}$) in a manner proportional to $\hat{Q}_{\mathrm{md}}$ (not $\hat{A}$).
The structure of MD in $Q_{\mathrm{md}}$ therefore registers information concerning the back reaction on the system. This is measured with RWM, thereby reducing the  second uncertainty (which results in PPS-mixtures).
The new RWM 
introduced in \S \ref{rwm} 
uses components of both WM criterion.  The first, minimizing $\lambda\Delta Q_{\mathrm{md}}$, is introduced in greater depth in the next section and the second, minimizing $\Delta A_w$
is introduced in greater depth in \S \ref{wvavgop}.

\label{errors}

\subsection{\bf  Statistical Weak Measurements (SWM); $\lambda>\frac{1}{\sqrt{N}A_w}$}

\label{wvstat}

To make RWM more ``Gedanken-practical,"
we will consider Stern-Gerlach (SG) measurements in different contexts throughout this article where $\hat{A}$ will be a spin component (i.e. $\hat{\sigma}_{\xi}$) and $Q_{\mathrm{md}}$ will be the translational coordinate of the particle in the same direction, $\vec{\xi}$ as the spin component.
Having the particles themselves serve as MDs with the information about the measurement stored in a degree of freedom ($P_{\mathrm{md}}^{\xi}$) separate from the pre- or post-selection (so that there is no coupling  between  the  variable  in which the result of the measurement is stored and the  post-selection device)
provides the easiest way
to pick out only those MDs which are associated with those systems that satisfied the proper post-selection criteria:  the post-selection of the particles then also selects out the relevant MDs. 

 Suppose we pre-select a spin-1/2 system with $|\hat{\sigma}_x=+1\ra=\vert\!\!\uparrow_x\!\rangle $ at time $t_{\mathrm{in}}$.
{To do this, we filter $\hat{\sigma}_x=-1$ out of the beam by applying an inhomogeneous magnetic field, described by $H_{\mathrm{int}} =-\mu\hat{\sigma}_x \vec{B}_x$ where  $\vec{B}_x=\vec{x}B_0$.  The force on the particle $\frac{dB_x}{dt}=\mu B_0 \hat{\sigma}_x$ induces a change in momentum proportional to the time $T$ that the particle spends in the field, i.e. $\delta P_{\mathrm{md}}^x= \mu B_0\hat{\sigma}_x T$.  Since the particle is constrained to be in a region $\Delta x< D$ (with $D$ the size of the Stern-Gerlach opening), the initial uncertainty in the momentum must be $\Delta P_{\mathrm{md}}^x > \frac{1}{D}$. For this pre-selection measurement to create a  distinguishable split between  $\hat{\sigma}_x=+1$ and $\hat{\sigma}_x=-1$, the shift induced in the momentum by the inhomogeneous magnetic field must be greater than the uncertainty in the momentum, i.e. $\delta P_{\mathrm{md}}^x > \Delta P_{\mathrm{md}}^x$.} 
\footnote{The filter only interacts with the component of the spin which is not transmitted, e.g. the $\sigma_x=-1$ component would receive a strong repulsive interaction (via a potential $\hat{\sigma}_xB_x -B_x$ using a homogenous $B_x$), while the $\sigma_x=+1$ component would not have any change in it's momentum.}
We then perform a similar procedure to post-select $|\hat{\sigma}_{\mathrm{y}}= +1\ra=\vert\uparrow_y\rangle$ at time $t_{\mathrm{fin}}$.

If we now consider SG measurements in the intermediate time $t$, $t_{\mathrm{in}}<t<t_{\mathrm{fin}}$  at an angle $\xi$ to the $x-y$ plane, then the spin $\hat{\sigma}_{\xi}$ can be determined from the deflection of the particle which is proportional to the impulse $\delta P_{\mathrm{md}}^{\xi}=\lambda\hat{\sigma}_{\xi}$
imparted to the particle 
due to the inhomogeneous magnetic field which has a linear gradient in the same 
direction $\vec{\xi}$ as the spin component which is to be determined.
Since the particle is free, the spin is conserved in time and thus a measurement of either $\hat{\sigma}_x$ or $\hat{\sigma}_y$ at $t$ will yield $+1$. 
  This is also evident from ABL: the probability to obtain $\hat{\sigma}_{\xi}= +1$ in the intermediate time if an ideal measurement is performed is  $Pr(\hat{\sigma}_{\xi}=+1)=\frac{1+\cos (\xi) +\sin (\xi) + \cos (\xi) \sin(\xi)}{1+\cos(\xi)\sin(\xi)}$.
We see that if $\xi=0^\circ$ (i.e. $\hat{\sigma}_x$) then the intermediate measurement will yield $\hat{\sigma}_{\mathrm{x}}= +1$ with certainty  and when  $\xi=90^\circ$ (i.e. $\hat{\sigma}_y$), then the intermediate measurement will again yield $\hat{\sigma}_{\mathrm{y}}= +1$ with certainty.
Consider measuring  the spin in a
direction $\hat{\xi}=45^{\circ}$:
\beq
\hat{\sigma}_{\xi}=\hat{\sigma}_x \cos 45^{\circ} +\hat{\sigma}_y\sin 45^{\circ}=\frac{\hat{\sigma}_x +\hat{\sigma}_y}{\sqrt{2}}
\label{spin45}
\eeq
From the results $Pr(\hat{\sigma}_{x}=+1)=1$ and $Pr(\hat{\sigma}_{y}=+1)=1$, we might wonder if we could simply  plug in both their values $\hat{\sigma}_{\mathrm{x}}= +1$ and $\hat{\sigma}_{\mathrm{y}}= +1$ into eq. \ref{spin45} and obtain $\hat{\sigma}_{\xi}=\frac{1+1}{\sqrt{2}}=\frac{2}{\sqrt{2}}=\sqrt{2}$.
Such a result would obviously be incorrect for an ideal measurement because the eigenvalues of any spin operator, including $\hat{\sigma}_{\xi}$, are $\pm 1$.
We can also see from ${\left(\frac{\sigma_x+\sigma_y}{\sqrt{2}}\right)}^2=\frac{\sigma_x^2+\sigma_y^2+\sigma_x\sigma_y+\sigma_y\sigma_x}{2}=\frac{1+1+0}{2}=1$
   but implementing the above argument, we expect ${\left(\frac{\sigma_x+\sigma_y}{\sqrt{2}}\right)}^2={\left(\frac{1+1}{\sqrt{2}}\right)}^2=2\neq 1$.
Performing this step of replacing $\hat{\sigma}_{\mathrm{x}}= +1$ {\it and} $\hat{\sigma}_{\mathrm{y}}= +1$ in eq. \ref{spin45} can only be done if $\hat{\sigma}_{\mathrm{x}}$ and $\hat{\sigma}_{\mathrm{y}}$ commute, which would allow both values to be simultaneously definite.  The probability statements are only simultaneously true if we do not perform $\hat{\sigma}_y$ before $\hat{\sigma}_x$, since this would destroy $\ket{\Psi_{\mathrm{in}}}=\vert\!\!\uparrow_x\!\rangle$.
So, in general, the finding that $\hat{\sigma}_{\mathrm{x}}= +1$ with certainty or $\hat{\sigma}_{\mathrm{y}}= +1$ with certainty in the pre- and post-selected ensemble only held when {\it one} of these two measurements was performed in the intermediate time, not both.  
The physical reason that a measurement of $\hat{\sigma}_{\xi}$
   doesn`t produce $\sqrt{2}$ is that  the particle is exposed to
   a magnetic field with a {\it strong} gradient in the $\xi=45^\circ$ direction, which causes the spin to revolve
   around this axis in an uncertain fashion.
In other words, the conditions for an ideal measurement $\delta P_{\mathrm{md}}^{\xi}=\lambda \hat{\sigma}_{\xi}\gg\Delta P_{\mathrm{md}}^{\xi}$ 
will also necessitate $\Delta Q_{\mathrm{md}}^{\xi} \gg \frac{1}{\lambda\hat{\sigma}_{\xi}}$ which will thereby create a back-reaction causing a precession in the spin such that $\Delta \Theta \gg 1$ (i.e. more than one revolution), thereby destroying the information that in the past we had $\hat{\sigma}_x=+1$,
   and in the future we will have $\hat{\sigma}_y=+1$.  

However, there is a sense in which both $Pr(\hat{\sigma}_{x}=+1)=1$ {\it and} $Pr(\hat{\sigma}_{y}=+1)=1$
are simultaneously relevant for measurements in the intermediate time:  if  
  the measurement of $\hat{\sigma}_x$ {\it and} $\hat{\sigma}_y$ is performed (i.e. when $\hat{\sigma}_{\xi}$ is measured) in such a way that measurement of one  does {\it not} disturb the other, which is precisely what occurs in a WM. 
For such a WM, the inhomogeneity in the magnetic field induces a shift in momentum which is less than the uncertainty 
$\delta P_{\mathrm{md}}^\xi < \Delta P_{\mathrm{md}}^\xi$ and thus a wave packet corresponding to $\frac{\hat{\sigma}_x+\hat{\sigma}_y}{\sqrt{2}}=1$ will be broadly overlapping with the wave packet corresponding to $\frac{\hat{\sigma}_x+\hat{\sigma}_y}{\sqrt{2}}=-1$ because the deflection $\delta Q_{\mathrm{md}}^{\xi} \propto \langle \vec{\sigma}_{\xi}\cdot \nabla  B\rangle$ will not be discernable from the noise, i.e. $\delta Q_{\mathrm{md}}^{\xi}\ll \Delta Q_{\mathrm{md}}^{\xi}$ (where $\Delta Q_{\mathrm{md}}^{\xi}$ is the dispersion in the particle beam). Thus, it cannot be determined whether any individual particle corresponds to $\frac{\hat{\sigma}_x+\hat{\sigma}_y}{\sqrt{2}}=\pm 1$.  
Furthermore, since the MD is quite
imprecise, we cannot say whether the
distribution of results in the pointer was due to the original uncertainty in the pointer $\lambda \Delta Q_{\mathrm{md}}$,
or due to the distribution of the observable of the system being measured, $\Delta A$.  

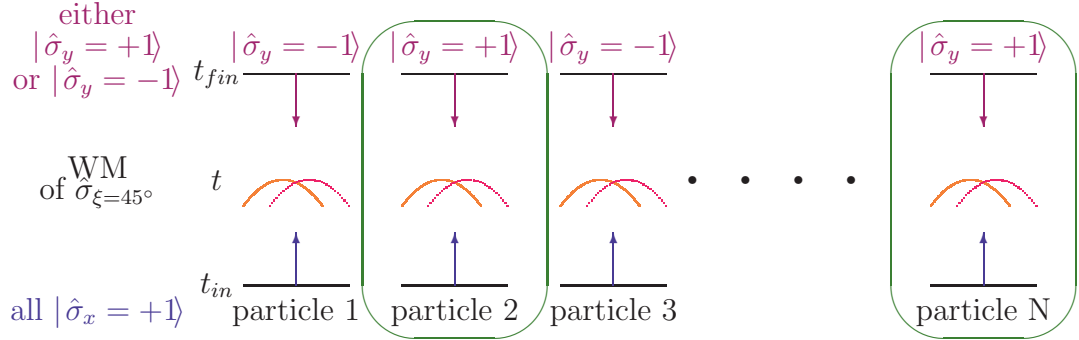
\begin{figure}[h] 
\vskip 1cm
\begin{picture}(400,90)(0,0)
\put(90,10){\line(1,0){40}}
\put(90,90){\line(1,0){40}}
\color{BlueViolet}
\put(35,0){\makebox(0,0){all $\mid\!\hat{\sigma}_x =+1\!\rangle$}}
\put(110,10){\vector(0,1){20}}
\color{RedViolet}
\put(110,90){\vector(0,-1){20}}
\put(35,113){\makebox(0,0){either}}
\put(35,100){\makebox(0,0){$\mid\!\hat{\sigma}_y =+1\!\rangle$}}
\put(35,87){\makebox(0,0){or $\mid\!\hat{\sigma}_y =-1\!\rangle$}}

\put(110,100){\makebox(0,0){$\mid\!\hat{\sigma}_y =-1\!\rangle$}}

\put(170,100){\makebox(0,0){$\mid\!\hat{\sigma}_y =+1\!\rangle$}}

\put(230,100){\makebox(0,0){$\mid\!\hat{\sigma}_y =-1\!\rangle$}}

\put(370,100){\makebox(0,0){$\mid\!\hat{\sigma}_y =+1\!\rangle$}}

\color{OliveGreen}

\put(170,50){\oval(70,120)}
\put(370,50){\oval(70,120)}

\color{Black}

\put(110,0){\makebox(0,0){particle 1}}
\put(150,10){\line(1,0){40}}
\put(150,90){\line(1,0){40}}
\color{BlueViolet}

\put(170,10){\vector(0,1){20}}
\color{RedViolet}

\put(170,90){\vector(0,-1){20}}

\color{Black}
\put(170,0){\makebox(0,0){particle 2}}
\put(210,10){\line(1,0){40}}
\put(210,90){\line(1,0){40}}
\color{BlueViolet}

\put(230,10){\vector(0,1){20}}
\color{RedViolet}

\put(230,90){\vector(0,-1){20}}

\color{Black}
\put(230,0){\makebox(0,0){particle 3}}

\put(260,50){\circle*{3}}
\put(280,50){\circle*{3}}
\put(300,50){\circle*{3}}
\put(320,50){\circle*{3}}
\put(350,10){\line(1,0){40}}
\put(350,90){\line(1,0){40}}
\color{BlueViolet}

\put(370,10){\vector(0,1){20}}
\color{RedViolet}

\put(370,90){\vector(0,-1){20}}
\color{Black}
\put(370,0){\makebox(0,0){particle N}}
\put(80,10){\makebox(0,0){$t_{in}$}}
\put(80,90){\makebox(0,0){$t_{fin}$}}
\put(80,50){\makebox(0,0){$t$}}
\put(35,55){\makebox(0,0){WM}}
\put(35,45){\makebox(0,0){of $\hat{\sigma}_{\xi=45^\circ}$}}

\color{Orange}
\bezier{500}(90,40)(105,60)(120,40)

\color{WildStrawberry}
\bezier{35}(100,40)(115,60)(130,40)

\color{Orange}
\bezier{500}(150,40)(165,60)(180,40)
\color{WildStrawberry}
\bezier{35}(160,40)(175,60)(190,40)

\color{Orange}
\bezier{500}(210,40)(225,60)(240,40)
\color{WildStrawberry}
\bezier{35}(220,40)(235,60)(250,40)

\color{Orange}
\bezier{500}(350,40)(365,60)(380,40)
\color{WildStrawberry}
\bezier{35}(360,40)(375,60)(390,40)

\end{picture}

\caption[Complete correlation between N particles]{Statistical WM ensemble.}
{\small }
\label{multparticlespre}
\end{figure}

Nevertheless, with statistical weak measurements (SWM), 
the WV can be obtained robustly in a statistical sense 
from the mean reading of many separate pointers (see fig. \ref{multparticlespre}). For example, an ensemble of $N$ separate systems and $N$ separate MDs are used (which again for SG are independent degrees-of-freedom of the same particle being measured). For {\bf each} individual system, in between it's pre- and post-selection, one of the $N$ MDs weakly measures the observable $\hat{A}$ of this single system and the outcome of this measurement is individually recorded. This is repeated for each of the $N$ different systems, each with a different MD. After the post-selection, the subset of those MDs which were associated with those systems which satisfied the post-selection criterion are collected out of the larger set of all possible post-selections (if all post-selections were included, then the decomposition, eq. \ref{expweak} would be reproduced).  While the WM was performed during the intermediate time $t$, we will obtain the same outcome and a simplified analysis if the readout of the WM MD is delayed until after the post-selection. 
A statistical analysis is then performed on the results of only those MDs associated with the proper post-selection and an average is manually calculated.  
This statistical procedure reduces the uncertainty in the mean position  by $\frac{1}{\sqrt{N}}$,  thereby allowing for a more precise calculation of $\hat{A}_w$ ~\cite{av}.
When thus correlated with the post-selection, the measurement result (which was confirmed experimentally for an analogous observable, the polarization~\cite{RSH}) is:
\begin{equation}
(\hat{\sigma}_{\xi=45^\circ})_{\mathrm{w}} =\frac{\langle \uparrow_y\vert \frac{\hat{\sigma}_y+\hat{\sigma}_x}{\sqrt{2}} 
\vert\!\uparrow_x
\rangle}{\langle{\uparrow_y}\vert{\uparrow_x}\rangle}=
\frac{{\left\{\langle \uparrow_y\vert \hat{\sigma}_y\right\}+\left\{\hat{\sigma}_x\vert\!\uparrow_x
\rangle\right\}}}  {\sqrt{2}\langle{\uparrow_y}\vert{\uparrow_x}\rangle}=\frac{\langle \uparrow_y\vert 1+1 
\vert\!\uparrow_x
\rangle}{\sqrt{2}\langle{\uparrow_y}\vert{\uparrow_x}\rangle}= \sqrt{2}
\end{equation}
For an individual spin, the component of spin
$\hat{\sigma}_{\hat{\xi}}$ is an eigenvalue, $\pm 1$, but the WV
$(\hat{\sigma}_{\hat{\xi}})_w=\sqrt{2}$ is $\sqrt{2}$ times
bigger,
 (i.e. lies outside the range of 
eigenvalues of ${\bf \hat{\sigma} \cdot n}$) and is thus called an ``eccentric weak value" (EWV)\footnote{WVs even further outside the eigenvalue spectrum which therefore offer the possibility of even greater amplification can be obtained by post-selecting states which are more anti-parallel to the pre-selection: e.g. if we post-select the $+1$ eigenstate of $(\cos\alpha)\sigma_x + (\sin\alpha)\sigma_z$, then $(\hat{\sigma}_z)_{\mathrm{w}}=\lambda\tan \frac{\alpha}{2}$.}.  In this case, we can use EWVs to amplify either the magnetic moment $\mu$ or the gradient in the field.

If we now consider a finite sample of particles, we ask what are the minimum number of particles necessary in order to distinguish the WV ``signal"  $(\hat{\sigma}_{\xi})_{\mathrm{w}}$ from the noise, i.e. such that the total momentum shift is greater than the total deviation $\delta P_{\mathrm{md}}\upn\gg\Delta P_{\mathrm{md}}\upn$. 
E.g. suppose the momentum is deposited onto a photographic plate after the post-selection.  From the WM interaction with each of the $N$ particles, the total momentum deposited on the plate will be 
$\delta P_{\mathrm{md}}\upn=N\lambda (\hat{\sigma}_{\xi})_{\mathrm{w}}$.  Now, for simplicity, we set $\Delta P_{\mathrm{md}}^{\xi}=1$ and thus the dispersion is also
$\tilde{\Delta} P_{\mathrm{md}}^{\xi}=(\Delta P_{\mathrm{md}}^{\xi})^2=1$. 
The total dispersion 
is $\tilde{\Delta} P_{\mathrm{md}}\upn=\sum \tilde{\Delta} P_{\mathrm{md}}^{\xi}=N\tilde{\Delta} P_{\mathrm{md}}^{\xi}$.  However, the standard deviation 
$\Delta P_{\mathrm{md}}\upn=\sqrt{\tilde{\Delta} P_{\mathrm{md}}\upn}$ and therefore, in order to distinguish the outcomes,  
$\delta P_{\mathrm{md}}\upn=N\delta P_{\mathrm{md}}^{\xi}=N\lambda({\sigma_{\xi}})_w\gg\Delta P_{\mathrm{md}}\upn=\sqrt{N}{\Delta} P_{\mathrm{md}}^{\xi}=\sqrt{N}$.
Therefore, the center of this distribution can be determined by using an ensemble of $N$ particles, where $N>\left\{\frac{\Delta P_{\mathrm{md}}^{\xi}}{\delta P_{\mathrm{md}}^{\xi}} \right\}^2$.

What, generally, are the limitations of this method, SWM? 
Suppose $N=20$ and that in order to satisfy $\delta P_{\mathrm{md}}\upn\gg \Delta P_{\mathrm{md}}\upn$ we are content with $\delta P_{\mathrm{md}}\upn\geq 2\Delta P_{\mathrm{md}}\upn$.  In this case, we must increase the coupling $\lambda$ to overcome the fluctuations, so from $\delta P_{\mathrm{md}}\upn\geq 2\Delta P_{\mathrm{md}}\upn$ and with $(\hat{\sigma}_{\xi=45^\circ})_{\mathrm{w}} = \sqrt{2}$, we have $\lambda(\hat{\sigma}_{\hat{\xi}})_w N\geq 2\sqrt{N}$ or $\lambda \geq \frac{2}{(\hat{\sigma}_{\xi})_{\mathrm{w}}}\frac{\sqrt{N}}{N}=\sqrt{\frac{1}{10}}$.  However, with the SWM requirement of minimizing $\lambda \Delta Q_{\mathrm{md}}$, a WM interaction strength of $\lambda=\sqrt{\frac{1}{10}}$ is too large  for a valid WM.  When such a WM is attempted, $\lambda \Delta Q_{\mathrm{md}}$ will have a significant back-reaction and will thereby create uncertainty in the boundary conditions.  
The reason we are interested in the minimum of $\lambda$ rather than its maximum is that we need  to obtain an EWV (i.e. outside the eigenvalue spectrum), rather than an ordinary WV, in order to implement our amplification scheme.   Now that we are dealing with finite samples, there will of necessity be a back-reaction on the system due to the WM.
If the back-reaction  rotates the pre- or post-selection by too much (e.g. by  $\pi$) then we will not obtain an EWV but rather an ordinary WV.  In addition, this will rotate the pre-selected state by an uncertain angle and therefore we will not know the WV with certainty as was required to implement the amplification scheme because we don't know whether the outcome we obtained is related more to $\mu$ or to $A_w$.  
In other words, up to this point we have argued that as a limiting process (i.e. when there is no back-reaction), we can measure the WV that would have been there even if no measurement was actually made.  However, the back-reaction invalidates this approach.
So, to  control the rotation of the pre- or post-selection
we need to control $\lambda \Delta \hat{Q}$ which is responsible for the back-reaction.
Thus, while we are certainly able to measure $A_w$ up to $O(\frac{1}{\sqrt{N}})$, we do not know which $A_w$ we are actually measuring.

This limitation of SWM (i.e. $N>\frac{1}{(\lambda(\hat{\sigma}_{\xi})_{\mathrm{w}})^2}$, or alternatively $\lambda>\frac{1}{\sqrt{N}(\hat{\sigma}_{\xi})_{\mathrm{w}}}$) is altered by RWM which reduces the uncertainty in $A_w$ that is created when the back-reaction of MD changes the post-selection in an uncertain manner.
\subsection{\bf Single Trial Weak Measurements (STWM)}
\label{wvavgop}

The limitation of SWMs, $\lambda>\frac{1}{\sqrt{N}(\hat{\sigma}_{\xi})_{\mathrm{w}}}$, is significant when dealing with finite sample sizes.  We can do better, however, 
by measuring {\it collective observables} 
In addition, this
allows us to measure all WVs with great precision
in one single (though previously thought to be rare) experiment.   
We consider again our theorem: $\hat{A} |\Psi \rangle = \la\hat{A}\ra \ket{\Psi}  + \Delta A \ket{\Psi_\perp}$~\cite{avg}.
We can also {\it measure} this property with no reference to statistics by applying this identity to a composite, $N$-particle state (which can also be viewed as a single system such as a large spin)
$\ket{\Psi\upn} =\ket{\Psi}_1\ket{\Psi}_2....\ket{\Psi}_N$
and using 
a ``collective operator," $\hat{A}\upn\equiv \frac{1}{N} \sum_{\mathrm{i=1}}^{N} \hat{A}_i$ (where  $\hat{A}_i $ is the same operator $\hat{A}$ acting on the $i$-th particle).
Using this, we are able to obtain information on $\la\hat{A}\ra$  without causing a collapse and thus without using a statistical approach 
because any product state
  $|\Psi \upn\rangle$ becomes an eigenstate of the operator $\hat{A}\upn$. 
To see this, consider~\cite{av,ah1990} $
{\hat{A}}\upn\ket{\Psi\upn}$:  
\beq
\hat{A}\upn\ket{\Psi\upn}  = \frac{1}{N}\left[ N \la\hat{A}\ra\ket{\Psi\upn} + \Delta A \sum_i
|\Psi\upn_\perp(i) \rangle \right]
\label{avgop}
\eeq
where $\la\hat{A}\ra$ is the average for any one particle and the states $|\Psi\upn_\perp(i) \rangle$ are mutually orthogonal and are given by
$|\Psi\upn_\perp(i) \rangle = \ket{\Psi}_1\ket{\Psi}_2...\ket{\Psi_\perp}_i...\ket{\Psi}_N$.
That is, the $i$th state has particle $i$ changed to an orthogonal state and all the other particles
remain in the same state.  If we further define a normalized state
 $|\Psi\upn_{\perp} \rangle = \sum_{i}\frac{1}{\sqrt{N}}|\Psi\upn_\perp(i) \rangle$ 
then the last term of eq. \ref{avgop} is $\frac{\Delta A}{\sqrt{N}}|\Psi\upn_{\perp} \rangle$ and it's size is $|\frac{\Delta A}{\sqrt{N}}
|\Psi\upn_{\perp} \rangle|^2\propto\frac{1}{N}\rightarrow 0$.
Therefore, $|\Psi\upn\rangle$ becomes an eigenstate of $\hat{A}\upn$, with value
$\overline{A}$, as $\hat{N} \rightarrow \infty$ (the second term decreases as $O(N^{-1/2})$ even if the particles are not all in the same state, as long as the composite
$N$-particle state is a product state).

We shall now consider a similar setup as used in \S \ref{wvstat}.  We perform  a WM of the collective observable in the $45^{\circ}$ angle to the $x-y$ plane of $\hat{\sigma}_\xi\upn \equiv \frac{1}{N}
\sum_{\mathrm{i=1}}^{N} \hat{\sigma}_\xi^i$~\cite{aacv}. Using 
$H_{\mathrm{int}} = -{{\lambda  \delta(t)}\over N}  \hat{Q}_{\mathrm{md}} \sum_{\mathrm{i=1}}^N  \hat{\sigma}^i_\xi$, a particular pre-selection of $|{\uparrow_x} \rangle$ (i.e. $|\Psi_{\mathrm{in}}\upn\rangle = \prod_{\mathrm{j=1}}^N |{\uparrow_x} \rangle_j$) and post-selection
 $|{\uparrow_y}\rangle$
(i.e. 
$\langle\Psi_{\mathrm{fin}}\upn|  = \prod_{\mathrm{k=1}}^N \langle{\uparrow_y}|_k=\prod_{\mathrm{n=1}}^N \left\{|{\uparrow_z} \rangle_n+i|{\downarrow_z}
\rangle_n\right\}$), we will show that the pointer is robustly shifted by the
the same WV obtained in \S \ref{wvstat}, i.e. $\sqrt{2}$:
\beq
  \label{wvC}
  (\hat{\sigma}_\xi)_{\mathrm{w}} = {{\prod_{k=1}^N \langle{\uparrow_y}|_k ~ \sum_{\mathrm{i=1}}^N
\left\{\hat{\sigma}^i_x + \hat{\sigma}^i_y\right\} ~ \prod_{\mathrm{j=1}}^N |{\uparrow_x} \rangle_j}
\over { \sqrt 2 ~ N(\langle{\uparrow_y} |{\uparrow_x} \rangle)^N}}=
\sqrt2 \pm O(\frac{1}{\sqrt N}).
\label{wvlargespin}
\eeq

Using $[\hat{\sigma}_i\upn,\hat{\sigma}_j\upn]=\frac{2i}{N^2}\varepsilon_{ijk}\sum_n\hat{\sigma}_k^{n}$, we will see that with $N\rightarrow\infty$, all these operators commute:
   \begin{displaymath}
      |\langle i[\hat{\sigma}_i\upn,\hat{\sigma}_j\upn]\rangle|=|\langle i\frac{2i}{N^2}\varepsilon_{ijk}\sum_n\hat{\sigma}_k^{n} \rangle|=
      |\frac{-2}{N^2}\varepsilon_{ijk}\sum_n\langle\hat{\sigma}_k^{n}\rangle|\leq\frac{2}{N^2}N=O(\frac{1}{N})
   \end{displaymath}
we see that for any given state and sufficiently large $N$, we may neglect
   the fact that these operators do not commute.
   In addition, for sufficiently large $N$, we may measure $\hat{\sigma}_x\upn$ (or $\hat{\sigma}_y\upn$) and the probability for a collapse can be made arbitrarily small.
Using again the theorem, we have $\hat{\sigma}_x\upn|\Psi_{\mathrm{in}}\upn\rangle=\langle\hat{\sigma}_x\upn\rangle|\Psi_{\mathrm{in}}\upn\rangle+\Delta\hat{\sigma}_x\upn|\Psi_{\mathrm{in\perp}}\upn\rangle$.
   We will now show that as $N\rightarrow\infty$, ${\Delta\hat{\sigma}_x\upn}\longrightarrow 0$.  In addition, from this we conclude that at the limit, $|\Psi_{\mathrm{in}}\upn\rangle$ is an eigenstate of $\hat{\sigma}_x\upn$
   which means that not even one of the spins will collapse.   To calculate $\Delta \hat{\sigma}_x\upn$ we need $\langle\hat{\sigma}_x\upn\rangle=\langle\frac{1}{N}\sum_{n=1}^N\hat{\sigma}_x^{n}\rangle=\frac{1}{N}\sum_{n=1}^N\langle\hat{\sigma}_x^{n}\rangle$ and 
   because the spin states are identical $\langle\hat{\sigma}_x\upn\rangle=\langle\hat{\sigma}_x^{1}\rangle$.  We will also need:

   \begin{displaymath}
      \langle(\hat{\sigma}_x\upn)^2\rangle=\bigg<\left(\frac{1}{N}\sum_{n=1}^N\hat{\sigma}_x^{n}\right)\left(\frac{1}{N}\sum_{m=1}^N\hat{\sigma}_x^{m}\right)\bigg>
      =\frac{1}{N^2}\left(N\langle(\hat{\sigma}^{1}_x)^2\rangle + \sum_{n=1}^N\sum_{m=1 \atop {m\neq n}}^N\langle\hat{\sigma}_x^{n}\hat{\sigma}_x^{m}\rangle\right)
   \end{displaymath}

\noindent now since $\hat{\sigma}_x^{n}$ and $\hat{\sigma}_x^{m}$ operate in different spaces,
   \begin{displaymath}
      \langle(\hat{\sigma}_x\upn)^2\rangle=\frac{1}{N^2}\left(N\langle(\hat{\sigma}^1_x)^2\rangle+N(N-1){\langle{\hat{\sigma}^{1}}_x\rangle}^2\right)=
      {\langle{\hat{\sigma}^{1}}_x\rangle}^2+\frac{1}{N}(\langle({\hat{\sigma}^{1}}_x)^2\rangle-{\langle{\hat{\sigma}^{1}}_x\rangle}^2)
   \end{displaymath} 
   \begin{equation}
      \mathrm{and\,\,\, finally,\,\,\,\,\,\,\,\,\,\,\,\,\,\,\,\,\,\,\,\,\,\,\,\,\,\,\,\,\,\,\,\,\,\,\,\,} {\Delta\hat{\sigma}_x\upn}^2=\langle(\hat{\sigma}_x\upn)^2\rangle-{\langle\hat{\sigma}_x\upn\rangle}^2=
      \frac{1}{N}(\langle(\hat{\sigma}^1_x)^2\rangle-{\langle{\hat{\sigma}^{1}}_x\rangle}^2)=O(\frac{1}{N})
\label{deltaaw}
   \end{equation}
To obtain eq. \ref{wvlargespin}, let us now calculate the final state of MD after post-selection:
\beq
|\Phi_{\mathrm{in}}^{\mathrm{MD}}\ra=\prod_{j=1}^N \langle{\uparrow_y}|_j \exp\left\{{{\lambda}\over N}  \hat{Q}_{\mathrm{md}} \sum_{\mathrm{k=1}}^N  \hat{\sigma}^k_\xi\right\}  \prod_{i=1}^N|{\uparrow_x} \rangle_i |\Phi_{\mathrm{in}}^{\mathrm{MD}}\ra
\label{bigspinb}
\eeq
Since the spins do not interact with each other, we can calculate one of the products and take the result to the $Nth$ power and 
eq. \ref{bigspinb} can be re-written:
\beq
|\Phi_{\mathrm{in}}^{\mathrm{MD}}\ra=\prod_{j=1}^N \langle{\uparrow_y}|_j \exp\left\{{{\lambda}\over N}  \hat{Q}_{\mathrm{md}}  \hat{\sigma}^j_\xi\right\} |{\uparrow_x} \rangle_j |\Phi_{\mathrm{in}}^{\mathrm{MD}}\ra=\left\{\langle{\uparrow_y}| \exp\left\{{{\lambda}\over N}  \hat{Q}_{\mathrm{md}}  \hat{\sigma}_\xi\right\} |{\uparrow_x} \rangle\right\}^N\!\!\!\! |\Phi_{\mathrm{in}}^{\mathrm{MD}}\ra
\label{bigspinb}
\eeq
Using the following identity 
   $\exp\left\{{i\alpha\hat{\sigma}_{\vec{n}}}\right\}=\cos\alpha+i\hat{\sigma}_{\vec{n}}\sin\alpha$~\cite{proof}, 
this becomes:
\begin{eqnarray}
\Phi_{\mathrm{fin}}^{\mathrm{MD}}&=&\left\{\langle{\uparrow_y}| \left[\cos \frac{{\lambda} \hat{Q}_{\mathrm{md}}}{N}-i\hat{\sigma}_\xi\sin \frac{{\lambda} \hat{Q}_{\mathrm{md}}}{N}\right]  |{\uparrow_x} \rangle\right\}^N |\Phi_{\mathrm{in}}^{\mathrm{MD}}\ra\nonumber\\
&=&
{\left[\langle{\uparrow_y}|{\uparrow_x} \rangle\right]^N }
\left\{\cos \frac{{\lambda} \hat{Q}_{\mathrm{md}}}{N}-i\alpha_w\sin \frac{{\lambda} \hat{Q}_{\mathrm{md}}}{N}\right\}^N  |\Phi_{\mathrm{in}}^{\mathrm{MD}}\ra
\label{bigspina}
\end{eqnarray}
where we have substituted $\alpha_w\equiv(\hat{\sigma}_\xi)_w=\weakv {\uparrow_y}{\hat{\sigma}_\xi}{\uparrow_x }$.    We consider only the second part (the first bracket, a number, can be neglected since it does not depend on $\hat{Q}$ and thus can only affect the normalization):
\beq
\Phi_{\mathrm{fin}}^{\mathrm{MD}}=
\left\{1-\frac{{\lambda}^2 (\hat{Q}_{\mathrm{md}})^2}{N^2} - \frac{i{\lambda} \alpha_w\hat{Q}_{\mathrm{md}}}{N} \right\}^N|\Phi_{\mathrm{in}}^{\mathrm{MD}}\ra\nonumber\\
\label{bigspinc}
\eeq
As $N\rightarrow\infty$, we use 
$(1+\frac{a}{N})^N=(1+\frac{a}{N})^{\frac{N}{a}a} \approx e^a$ and obtain $|\Phi_{\mathrm{fin}}^{\mathrm{MD}}\ra\approx e^{i\lambda\alpha_w  \hat{Q}_{\it md}^{(N)} } |\Phi_{\mathrm{in}}^{\mathrm{MD}}\ra$.
When projected onto $P_{\mathrm{md}}$, this results in the same shift $\sqrt{2}$ as the SWM example in \S \ref{wvstat}.
The {\it maximum} of $\lambda$ can be increased up to $\epsilon\sqrt{N}$ from the weakness condition and eq. \ref{deltaaw}, the weak uncertainty $\Delta (\hat{\sigma}_{\xi})_{\mathrm{w}}\sim\frac{1}{\sqrt{N}}$.
In this case, an individual spin is acted on by $U=\exp\left\{{{\lambda}\over N}  \hat{Q}_{\mathrm{md}}   \hat{\sigma}_\xi\right\}$, 
 which rotates the spin around the $\xi$-axis by an uncertain angle $\Delta \theta\approx \frac{\lambda\Delta {Q}_{\mathrm{md}}}{N}=\frac{\epsilon}{\sqrt{N}}$.  Thus, the probability that an individual spin is still in its original state is $1-\frac{\epsilon ^2}{N}$ and the probability that the entire $N$ spin system stays in it's original state is $\{1-\frac{\epsilon ^2}{N}\}^N\approx \exp\{-\epsilon^2\}\rightarrow 1$ for $\epsilon \ll 1$.
Thus, as $N \rightarrow \infty$ the intermediate WM will give $ (\hat{\sigma}_\xi)_{\mathrm{w}} = \weakv {\Psi_{\mathrm{fin}}}{ \hat{\sigma}_\xi}{\Psi_{\mathrm{in}}}\,=\sqrt{2}$ robustly since the shift in MD $\delta P_{\mathrm{md}}=N\lambda (\hat{\sigma}_\xi)w$ is greater than the uncertainty in MD, $\Delta P_{\mathrm{md}}=1$.
A single experiment is now sufficient to
determine the WV with great precision and there is no longer
any need to average over results obtained in multiple experiments as we did in the previous section. 
Therefore, if we repeat the experiment with different MDs, then each MD will
show the very same WVs, up to an insignificant spread of $\frac{1}{\sqrt
{ N}}$ (assuming we obtain the particular, rare, post-selection).  Therefore,  the information from {\it both} boundary conditions, i.e. 
$|\Psi_{\mathrm{in}}\rangle = \prod_{\mathrm{i=1}}^N |{\uparrow_x} \rangle_i$
 and 
$\langle\Psi_{\mathrm{fin}}|  = \prod_{\mathrm{i=1}}^N \langle{\uparrow_y}|_i$, describes the entire interval of time between pre- and post-selection (for plots see \cite{vaidman,Unruh2}).

However, as pointed out in the beginning of this section, STWM has a major advantage over SWM  in the {\it minimum required} values for $\lambda$.
There is no difference in the momentum shifts for SWM and STWM as both cases are $\delta P_{\mathrm{md}}\upn=N(\hat{\sigma}_{\xi})_{\mathrm{w}}\lambda$.
However, for STWM, $\Delta P_{\mathrm{md}}\upn\sim 1$, whereas for SWM, $\Delta P_{\mathrm{md}}\upn\sim \sqrt{N}$.  To obtain $\delta P_{\mathrm{md}}\upn\gg \Delta P_{\mathrm{md}}\upn$, the {\it minimum} value for the coupling constant $\lambda$ for STWM can be smaller ($\lambda > \frac{2}{N(\hat{\sigma}_{\xi})_{\mathrm{w}}}$) by a factor of $\frac{1}{\sqrt{N}}$ than  for SWM ($\lambda > \frac{2}{\sqrt{N}(\hat{\sigma}_{\xi})_{\mathrm{w}}}$).  
There is thus a regime of $\lambda$  for which  the amplification scheme is invalid for individual particles, but for which it is valid for collective observables therefore substantially increasing the utility of STWM.

However, STWM has a major short-coming that is resolved by RWM.
While STWM is a valuable Gedanken experiment, 
the probability 
$|\langle\Psi_{\mathrm{fin}}|\Psi_{\mathrm{in}}\rangle|^{2N}$ for all $N$   particles to end up  in the same final state $|\Psi_{\mathrm{fin}} 
\rangle$ becomes exponentially small.
With the particular STWM considered in this section, we have $N$ particles pre-selected with $\hat{\sigma}_x =1$, a WM of
$\hat{\sigma}_{45^{\circ}}$ (which
doesn't significantly disturb
the spins which are thus still in the state $\hat{\sigma}_x =1$ after the $\hat{\sigma}_{45^{\circ}}$ measurement) and followed by a post-selection in the y-direction.  The probability to obtain $\hat{\sigma}_y =1$ is $1/2$ and thus the total probability of finding all $N$
spins with $\hat{\sigma}_y =1$ is an exponentially small $2^{-N}$.  

On the other hand,  SWM requires a much smaller sample.  For the particular pre- and post-selection used in \S \ref{wvstat},  approximately $N/2$ out of $N$ pre-selected particles will satisfy the post-selection criterion  
 and thus this result is not a rare outcome, making it much more attractive for our amplification scheme.  

\section{\textcolor{black}{\bf  Robust Weak Measurements on Finite Samples (RWM)}}
\label{rwm}

The RWM introduced in this section shares positive attributes of both the ``statistical" (\S \ref{wvstat}) and ``single-trial" (\S \ref{wvavgop}) approaches and seeks to minimize the 2 uncertainties discussed in \S \ref{der2v} which resulted from a finite sample size.
\begin{enumerate}
\item 
We keep a significant benefit from the first SWM approach, namely the PPS-ensemble necessary to obtain EWVs was not rare.
However, the SWM has a major disadvantage for finite samples, namely in order to distinguish the WV from the noise, a relatively large coupling constant is required, $\lambda > \frac{2}{\sqrt{N}(\hat{\sigma}_{\xi})_{\mathrm{w}}}$ which can cause a significant back-reaction on the system and therefore increase the second uncertainty in $A_w$, creating a mixture of WVs.

\item 
In the SWM approach, we  measured each individual shift $\delta P_{\mathrm{md}}$ and obtained the same total shift $\delta P_{\mathrm{md}}\upn=N(\hat{\sigma}_{\xi})_{\mathrm{w}}\lambda$ as was obtained for STWM,  which did not involve a measurement of ``individual" shifts.  
Therefore,  with SWM we are in a sense doing too much because there is additional information which can be obtained (namely the relative positions which commute with the total momentum) if we wait to make the measurement of $\delta P_{\mathrm{md}}\upn$ as in STWM.
We can then use the relative positions to correct for the disturbance caused to the system which resulted from a stronger measurement interaction which was required by the use of part of the SWM approach.  
\end{enumerate}

\begin{figure}[here]
\scalebox{.6}{\includegraphics{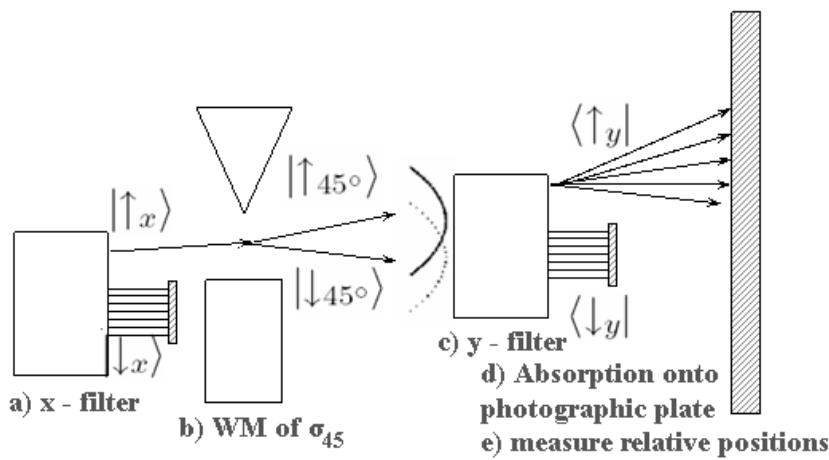}}
\caption[Stern Gerlach Apparatus]{RWM by Stern-Gerlach apparatus for weakly measuring $\hat{\sigma}_{45^{\circ}}$.}
\label{sgpre}
\vskip -.5cm
\end{figure}

Without loss of generality, we present the new RWM in the framework of SG measurements used in previous sections:
consider a large collection of particles where the MD is simply the position and momentum of those particles (see fig. \ref{sgpre}).  We perform the following a) filter out $|{\downarrow_x} \rangle$ at time $t_{\mathrm{in}}$; b) perform a WM of $\frac{\hat{\sigma}_x+\hat{\sigma}_y}{\sqrt{2}}$ at the intermediate time $t$ but wait until after performing the post-selection to read out the result of the sum of $\approx \frac{1}{2}$ of these interactions\footnote{While we present this Gedankenexperiment in the same spirit as the Einstein  Gedankenexperiment, we also recommend that a  WM of $\sigma_x$ uses a field $B_o(\sigma_x x-\sigma_y y)$ with a small width in $y$.  There is then little variation in the wavefunction in the y-direction.  The x-direction  would not be constrained and the wavefunction can freely vary in $x$.  With this method, only a shift in the $x$ direction would occur; the wavefunction in $y$ would always be left in the ground state because the force is too small to excite it.}; c) filter out  $\langle{\downarrow_y}|$ at time $t_{\mathrm{fin}}$; 
d) absorb the particles onto a photographic plate 
and measure the sum of momenta  $\sum_{\mathrm{i=1}}^{N} \hat{P}_{\mathrm{md}}^i$
(without measuring the individual $\hat{P}_{\mathrm{md}}^i$); this recording  will produce a definite shift 
by a WV; e) measure the relative positions to determine what the pre-selected and post-selected system the WM in step (b) was a measurement of. 
\vskip -.1cm

Let us consider again 
a particular pre-selection for each particle of $|{\uparrow_x} \rangle$ and post-selection in the
state $|{\uparrow_y}\rangle$, so $|\Psi_{\mathrm{in}}\rangle = \prod_{\mathrm{j=1}}^N |{\uparrow_x} \rangle_j$ 
 and  $|\Psi_{\mathrm{fin}}\rangle  = \prod_{\mathrm{j=1}}^N |{\uparrow_y}\rangle_j$. 
The WM interaction in step b)~\cite{aacv} is  described by an interaction Hamiltonian which couples $\hat{\sigma}_\xi$ with $\hat{Q}_{\mathrm{md}}$ of MD
i.e. $H_{\mathrm{int}} = \lambda  \delta(t)   \hat{Q}_{\mathrm{md}}^j\hat{\sigma}^j_\xi$ (where $\hat{\sigma}^j_\xi=\{ \frac{\hat{\sigma}_x+\hat{\sigma}_y}{\sqrt{2}}\}_j$).
This generates shifts in the individual momenta due to the WM interaction (as occurred in \S \ref{wvstat}).  However, unlike \S \ref{wvstat} there is only {\bf one irreversible recording} of the sum of these shifts (as occurred in \S \ref{wvavgop}), i.e. one irreversible recording of the total momentum of the $N$ particles which were deposited onto a single photographic plate, followed by $N$ measurements of the  positions (used to deduce the $N-1$ relative positions).  
This state of the photographic plate after its interaction with the $N$ particles shall be referred to as the ``final state of the MD".  
 After the WM interaction and post-selection  (but before the irreversible recording), 
the final state of MD is:
\beq
|\Phi^{\mathrm{fin}}_{\mathrm{md}}\ra=\prod_{\mathrm{j=1}}^N \langle{\uparrow_y}|_j \exp \lbrace {\mathrm{i}}\lambda \hat{Q}_{\mathrm{md}}^j \hat{\sigma}^j_\xi\rbrace  |{\uparrow_x} \rangle_j |\Phi^{\mathrm{in}}_{\mathrm{md}}\ra
\label{bigspin}
\eeq
Here we have set the coupling to each spin to be $ \int \lambda  \delta(t)dt=\lambda $
and without loss of generality have taken the initial state of MD as simply a Gaussian in the coordinate $\hat{Q}_{\mathrm{md}}^j$ of each particle, i.e. $(\Phi^{\mathrm{in}}_{\mathrm{md}})^j=\exp\left\{- {({{Q}_{\mathrm{md}}^j)^2}\over{4(\Delta {Q}_{\mathrm{md}}^j)^2}}\right\}$.
As will be seen later, it will prove useful to reformulate the MD observables in terms of two complementary, non-commuting, collective observables, a pointer corresponding to the sum of momenta, $\hat{P}_{\mathrm{md}}^{\mathrm{(N)}}$, and it's conjugate $\hat{Q}_{\mathrm{md}}^{\mathrm{(N)}}$ which generates shifts in the pointer $\hat{P}_{\mathrm{md}}^{\mathrm{(N)}}$: 
\begin{eqnarray}
\hat{P}_{\mathrm{md}}^{\mathrm{(N)}}&\equiv& \sum_{\mathrm{j=1}}^{N} \frac{\hat{P}_{\mathrm{md}}^j}{\sqrt{N}}\\
\hat{Q}_{\mathrm{md}}^{\mathrm{(N)}}&\equiv& \sum_{\mathrm{j=1}}^N \frac{\hat{Q}_{\mathrm{md}}^j}{\sqrt{N}}
\end{eqnarray}
  These definitions are particularly useful because
if the uncertainty in the individual $\hat{Q}_{\mathrm{md}}^j$'s is $\Delta \hat{Q}_{\mathrm{md}}^j\sim 1$, then the uncertainty in $\hat{Q}_{\mathrm{md}}^{\mathrm{(N)}}$ is also $\Delta \hat{Q}_{\mathrm{md}}^{\mathrm{(N)}}\sim 1$ due to $[\hat{Q}_{\mathrm{md}}^{\mathrm{(N)}}, \hat{P}_{\mathrm{md}}^{\mathrm{(N)}}]=1$ (The spread in $\sum_{\mathrm{j=1}}^N \hat{Q}_{\mathrm{md}}^j$ is $\sqrt{N}$ 
and thus $\frac{\sum_{\mathrm{j=1}}^N \hat{Q}_{\mathrm{md}}^j}{\sqrt{N}}\approx 1$).  Using

\begin{eqnarray}
- \sum_{\mathrm{j=1}}^N\frac{({Q}_{\mathrm{md}}^j)^2}{4(\Delta {Q}_{\mathrm{md}}^j)^2}&=&- \frac{1}{4(\Delta {Q}_{\mathrm{md}}^j)^2}\left[{\sum_{\mathrm{j=1}}^N{\lbrace{{Q}_{\mathrm{md}}^j}}- \frac{{Q}_{\mathrm{md}}^{\mathrm{(N)}}}{\sqrt{N}}\rbrace^2}+\sum_{\mathrm{j=1}}^N\lbrace  2{{Q}_{\mathrm{md}}^j}\frac{{Q}_{\mathrm{md}}^{\mathrm{(N)}}}{\sqrt{N}}-\frac{[{Q}_{\mathrm{md}}^{\mathrm{(N)}}]^2}{N}\rbrace\right]\nonumber\\
&=&- \frac{1}{4(\Delta {Q}_{\mathrm{md}}^j)^2}{\sum_{\mathrm{j=1}}^N{\lbrace{{Q}_{\mathrm{md}}^j}}- \frac{{Q}_{\mathrm{md}}^{\mathrm{(N)}}}{\sqrt{N}}\rbrace^2}
+[{Q}_{\mathrm{md}}^{\mathrm{(N)}}]^2
\end{eqnarray}
it will also be useful to re-write the wavefunction of MD as:
\beq
\prod_{\mathrm{j=1}}^N\exp\left\{- {({{Q}_{\mathrm{md}}^j)^2}\over{4(\Delta {Q}_{\mathrm{md}}^j)^2}}\right\}\rightarrow\exp{ \lbrace-{{({Q}_{\mathrm{md}}^{\mathrm{(N)}})^2}\over{4(\Delta {Q}_{\mathrm{md}}^{\mathrm{(N)}})^2}}\rbrace}\exp{ \lbrace-{{\sum_j({Q}_{\mathrm{md}}^j-\frac{{Q}_{\mathrm{md}}^{\mathrm{(N)}}}{\sqrt{N}})^2}\over{4(\Delta {Q}_{\mathrm{md}}^{\mathrm{(N)}})^2}}\rbrace}
\label{wfmd}
\eeq

We now show that measuring the relative positions provides corrections to the pre- or post-selection, thus giving a different WV for each particle, represented by $\tilde{\sigma}_{\mathrm{w}}^j$ (which thereby explains the utility of $\exp{ \lbrace-{{\sum_j({Q}_{\mathrm{md}}^j-\frac{{Q}_{\mathrm{md}}^{\mathrm{(N)}}}{\sqrt{N}})^2}\over{4(\Delta {Q}_{\mathrm{md}}^{\mathrm{(N)}})^2}}\rbrace}$).  


\subsection{\bf Use of relative positions}

\label{relposcorrect}

Besides the sum of momenta, we can also measure  the  $N-1$ relative positions (without disturbing the system),
\beq
\hat{x}_i=\hat{Q}_{\mathrm{md}}^i-\sum \frac{\hat{Q}_{\mathrm{md}}^n}{N}=\hat{Q}_{\mathrm{md}}^i-\frac{\hat{Q}_{\mathrm{md}}^{\mathrm{(N)}}}{\sqrt{N}}
\eeq
 This is because 
$[\hat{Q}_{\mathrm{md}}^i-\sum \frac{\hat{Q}_{\mathrm{md}}^n}{N},\sum \hat{P}_{\mathrm{md}}^i]=[\hat{x}_i,\hat{P}_{\mathrm{md}}^{\mathrm{(N)}}]=0$
which is easy to see because each pair of relative positions commutes with the sum of momenta, i.e. $[\hat{Q}_{\mathrm{md}}^i- \hat{Q}_{\mathrm{md}}^j,\sum_{\mathrm{n=1}}^{N} \hat{P}_{\mathrm{md}}^n]=[\hat{Q}_{\mathrm{md}}^i,\sum_{\mathrm{n=1}}^{N} \hat{P}_{\mathrm{md}}^n]-[\hat{Q}_{\mathrm{md}}^j,\sum_{\mathrm{n=1}}^{N} \hat{P}_{\mathrm{md}}^n]=i-i=0$, using $[\hat{Q}_{\mathrm{md}}^j,\sum_{\mathrm{n=1}}^{N} \hat{P}_{\mathrm{md}}^n]=[\hat{Q}_{\mathrm{md}}^j, \hat{P}_{\mathrm{md}}^j]=i$.  Furthermore $\hat{Q}_{\mathrm{md}}^i-\sum_{\mathrm{n=1}}^{N} \frac{\hat{Q}_{\mathrm{md}}^n}{N}=\sum_{\mathrm{n=1}}^{N} \frac{\hat{Q}_{\mathrm{md}}^i-\hat{Q}_{\mathrm{md}}^n}{N}$.

As a preparation to obtain both a measurement of the relative positions and of the total momenta  we re-write eq. \ref{bigspin} as: 
\begin{eqnarray}
\Phi^{\mathrm{fin}}_{\mathrm{md}}&=& \prod_{\mathrm{j=1}}^N \langle{\uparrow_y}|_j \exp \lbrace {\mathrm{i}}\lambda
{\{ \hat{Q}_{\mathrm{md}}^j-\sum_{\mathrm{n=1}}^N \frac{\hat{Q}_{\mathrm{md}}^n}{N}\}}
\hat{\sigma}^j_\xi\rbrace 
{\exp \lbrace {\mathrm{i}}\lambda  \sum_{\mathrm{n=1}}^N \frac{\hat{Q}_{\mathrm{md}}^n}{N} \hat{\sigma}^j_\xi \rbrace }
{|\uparrow_x} \rangle_j\nonumber\\
&\times & \exp{ \lbrace-{{({Q}_{\mathrm{md}}^{\mathrm{(N)}})^2}\over{4(\Delta {Q}_{\mathrm{md}}^{\mathrm{(N)}})^2}}\rbrace}\exp{ \lbrace-{{
\sum_j({Q}_{\mathrm{md}}^j-\frac{{Q}_{\mathrm{md}}^{\mathrm{(N)}}}{\sqrt{N}})^2}\over{4(\Delta {Q}_{\mathrm{md}}^{\mathrm{(N)}})^2}}\rbrace}\nonumber\\
&=&\prod_{\mathrm{j=1}}^N \langle{\uparrow_y}|_j \exp \lbrace {\mathrm{i}}\lambda \hat{x}_j \hat{\sigma}^j_\xi\rbrace \exp \lbrace {\mathrm{i}}
\lambda\frac{\hat{Q}_{\mathrm{md}}^{\mathrm{(N)}}}{\sqrt{N}} \hat{\sigma}^j_\xi \rbrace |{\uparrow_x} \rangle_j\nonumber\\
&\times & \exp{ \lbrace-{{({Q}_{\mathrm{md}}^{\mathrm{(N)}})^2}\over{4(\Delta {Q}_{\mathrm{md}}^{\mathrm{(N)}})^2}}\rbrace}\exp{ \lbrace-{{\sum_j({Q}_{\mathrm{md}}^j-\frac{{Q}_{\mathrm{md}}^{\mathrm{(N)}}}{\sqrt{N}})^2}\over{4(\Delta {Q}_{\mathrm{md}}^{\mathrm{(N)}})^2}}\rbrace}
\label{bigspin2}
\end{eqnarray}

How is this re-formulation of eq. \ref{bigspin} in terms of the relative positions $\hat{x}_j=\hat{Q}_{\mathrm{md}}^j-\sum \frac{\hat{Q}_{\mathrm{md}}^n}{N}$  helpful?  To see this, we'll consider eq. \ref{bigspin2} one particle at a time.  For the $jth$ particle, we can apply the first exponential of eq. \ref{bigspin2}, $\exp \lbrace i\lambda \hat{x}_j \hat{\sigma}^j_\xi\rbrace$, to either the pre-selected state $|{\uparrow_x} \rangle_j$  or to the post-selected state $\langle{\uparrow_y}|_j$ (since the 2 exponentials commute).  
What does this exponential do to   the pre- or post-selection? 
As  mentioned in \S \ref{der2v},
$\hat{Q}_{\mathrm{md}}^{\mathrm{(N)}}$ and $\hat{A}$ (or in this case $\hat{x}_j$ and $\hat{\sigma}^j_\xi$) have 2 inverse roles:  the back reaction on the {\it system} is generated by $\hat{\sigma}^j_\xi$ in a manner proportional to $\hat{x}_j$. 
However, the $\hat{x}_j$ can be measured exactly and can thus be replaced by a number.  Therefore,  $\exp \lbrace i\lambda \hat{x}_j \hat{\sigma}^j_\xi\rbrace$ simply 
 rotates the pre- or post-selected state about the axis $\xi $ by an angle given by $\lambda x_j$:
\beq
{\exp \lbrace {\mathrm{i}}\lambda \hat{x}_j 
\hat{\sigma}^j_\xi\rbrace}
|{\uparrow_x} \rangle_j\equiv |\Psi \rangle_j
\label{bigspin4}
\eeq
Thus, measurement of the relative positions allows us to definitely determine how much $\exp \lbrace {\mathrm{i}}\lambda \hat{x}_j 
\hat{\sigma}^j_\xi\rbrace$ rotates $|{\uparrow_x} \rangle_j$ (i.e. to $|\Psi \rangle_j$). 
Therefore, eq. \ref{bigspin4} acts as a correction to the ensemble:
instead of the original ensemble of pre-selected $|{\uparrow_x} \rangle$ and post-selected $\langle{\uparrow_y}|$ states, we will have a new  ensemble with  shifted pre- or post-selections. 

How could the relative positions be measured?  Procedurally, we first measure the momentum of the photographic plate after the $N$ particles have deposited their momentum.
When we subtract from this the initial  momentum of the photographic plate, then we can determine the shift in the sum of the momentum for the $N$ particles as a result of the WM interaction in a new way.  After the final measurement of $\hat{P}_{\mathrm{md}}^{\mathrm{(N)}}$, 
we then measure the $N$ individual positions (i.e. $\hat{Q}_{\mathrm{md}}^j$) of each particle that is deposited onto the photographic.
Now, measurement of $\hat{P}_{\mathrm{md}}^{\mathrm{(N)}}$
will disturb the individual $\hat{Q}_{\mathrm{md}}^j$'s but will not disturb the relative positions (since they commute with the total momenta).  
Therefore, even though the subsequent measurement of the $N$ $\hat{Q}_{\mathrm{md}}^j$'s will be un-related to the value of the $\hat{Q}_{\mathrm{md}}^j$'s during the WV,  
 we can {\it deduce} what the relative positions were at the time of the WM through the individual positions.
\footnote{
If the uncertainty in the individual $\hat{Q}_{\mathrm{md}}$'s is $\Delta \hat{Q}_{\mathrm{md}}\sim 1$, then the uncertainty in $\hat{Q}_{\mathrm{md}}^{\mathrm{(N)}}$ is also $\Delta Q_{\mathrm{md}}\sim 1$ (because the spread in $\sum\hat{Q}_{\mathrm{md}}$ is $\sqrt{N}$ and thus $\frac{\sum\hat{Q}_{\mathrm{md}}}{\sqrt{N}}\approx 1$).  Therefore, the spread in $\frac{\sum\hat{Q}_{\mathrm{md}}}{N}$ is negligible and thus $\hat{x}_i=\hat{Q}_{\mathrm{md}}^i-\sum \frac{\hat{Q}_{\mathrm{md}}^n}{N}$ also has the same uncertainty as $\Delta \hat{Q}_{\mathrm{md}}$.}
After substituting the single particle result 
eq. \ref{bigspin4}  for the $jth$ particle (i.e. using the rotated bra or ket), back into the $N$ particle eq. \ref{bigspin2}, we have:
\beq
\prod_{\mathrm{j=1}}^N \langle{\uparrow_y}|_j  \exp \lbrace {\mathrm{i}}\frac{\lambda}{\sqrt{N}}{\hat{Q}_{\mathrm{md}}^{\mathrm{(N)}}} \hat{\sigma}^j_\xi \rbrace |\Psi\rangle_j\exp\left\{-{{({Q}_{\mathrm{md}}^{\mathrm{(N)}})^2}\over{4(\Delta {Q}_{\mathrm{md}}^{\mathrm{(N)}})^2}}\right\}
\label{bigspin6}
\eeq
It is clear from eq. \ref{bigspin6} that when we look at the particles that are left unknown after using all the information (both relative positions and the total momenta) and consider them as the final total spin, then it is like a robust experiment but now the coupling to each spin is $\frac{\lambda}{\sqrt{N}}$ 
and thus the remaining effect of the coupling in the exponential will be small.

\subsection{\bf Proving the legitimacy of WVs for a new regime}
\label{newway}

After using these corrections, we can now prove the validity of the WV approximation. We will show that the final state of MD, i.e. of eq. \ref{bigspin6}  will be:
\beq
\Phi^{\mathrm{fin}}_{\mathrm{md}}= \exp \lbrace \frac{{\mathrm{i}}\lambda \hat{Q}_{\mathrm{md}}^{\mathrm{(N)}}}{\sqrt{N}} \sum_{\mathrm{j=1}}^N \tilde{\sigma}_{\mathrm{w}}^j  \rbrace\exp\lbrace-{{({Q}_{\mathrm{md}}^{\mathrm{(N)}})^2}\over{4(\Delta {Q}_{\mathrm{md}}^{\mathrm{(N)}})^2}}\rbrace
\label{finalwm}
\eeq
($\tilde{\sigma}_{\mathrm{w}}^j$ is the WV for the $jth$ particle
- a tilde will always refer to WVs calculated with rotated states)
When this is transformed back to the momentum 
representation (as was done in the WV approximation used in eq. \ref{mstate2}), then the momentum of MD is shifted by the WV; i.e. the change in $\hat{P}_{\mathrm{md}}^{\mathrm{(N)}}$ (the change in the sum of momentum $\sum_{\mathrm{i=1}}^{N}\hat{P}_{\mathrm{md}}^i$) is:
\beq
\delta \hat{P}_{\mathrm{md}}^{\mathrm{(N)}}=\delta \sum_{\mathrm{i=1}}^{N} \frac{\hat{P}_{\mathrm{md}}^i}{\sqrt{N}} = \frac{\lambda}{\sqrt{N}}\sum_{\mathrm{j=1}}^N \tilde{\sigma}_{\mathrm{w}}^j
\label{finprf}
\eeq
To prove the legitimacy of this WV calculation,  
we first assume for simplicity a small variance in the rotations so that each particle yields approximately the same WV, i.e. $\tilde{\sigma}_{\mathrm{w}}^j\equiv\bar{\alpha}_{\mathrm{w}}$, enabling us to re-write eq. \ref{finalwm} as:
\beq
\{cos \frac{\lambda \hat{Q}_{\mathrm{md}}^{\mathrm{(N)}}}{\sqrt{N}} +{\mathrm{i}}\bar{\alpha}_{\mathrm{w}} sin \frac{\lambda \hat{Q}_{\mathrm{md}}^{\mathrm{(N)}}}{\sqrt{N}}\}^N e^{-\frac{({Q}_{\mathrm{md}}^{\mathrm{(N)}})^2}{4(\Delta {Q}_{\mathrm{md}}^{\mathrm{(N)}})^2}}
\label{bigspin7}
\eeq
Now, in order to perform a valid WV calculation, this function needs to be peaked around $\hat{Q}_{\mathrm{md}}^{\mathrm{(N)}}=0$. As long as there are no regions in which the size of eq. \ref{bigspin7} (i.e. eq. \ref{bigspin8}) exceed the exponential of MD then it will be as if we are around $\hat{Q}_{\mathrm{md}}^{\mathrm{(N)}}=0$.   In other words, the legitimacy of the WV calculation can now be understood as a competition between the (scalar product) $A$ term and the (probability) $B$ term: 
\beq
|\Phi^{\mathrm{fin}}_{\mathrm{md}}|=\underbrace{\{1 + (\bar{\alpha}_{\mathrm{w}}^2 -1)sin^2 \frac{\lambda \hat{Q}_{\mathrm{md}}^{\mathrm{(N)}}}{\sqrt{N}}\}^{\frac{N}{2}}}_A \underbrace{\exp\left\{-\frac{({Q}_{\mathrm{md}}^{\mathrm{(N)}})^2}{2(\Delta {Q}_{\mathrm{md}}^{\mathrm{(N)}})^2}\right\}}_B
\label{bigspin8}
\eeq
If  the quantity eq. \ref{bigspin8} goes to $0$ for large $Q$ then the WV approximation is valid.
On the other hand, if the increase in  $A$ was not counter-balanced by the decline in $B$ then we could not restrict the WV approximation around $Q=0$ because it would be much more likely to be located around large $Q$.   Thus, the meaning of the new WV approximation presented here is that there is no other region in which the size of eq. \ref{bigspin7}, i.e. eq. \ref{bigspin8}, is significant, except around $Q=0$.
We now  ask what is the maximum value of $\lambda $ such that we still obtain a shift in the pointer by $\frac{\lambda}{\sqrt{N}}\sum_{\mathrm{j=1}}^N \bar{\alpha}_{\mathrm{w}}$? 
We will see that the constraint ${\lambda}  \ll 1$ as was required in \S \ref{wvstat} (and in the first article on WMs~\cite{aacv}) is unneccessary in order to obtain a valid WM.  In fact, with RWM the coupling to each individual spin just needs to be $\lambda \sim 1$.

\subsubsection{\bf  N$\rightarrow \infty$, $Q$ finite:} 


First we consider 
how large $\lambda$ can be for a legitimate WV in the regime $N\rightarrow 
\infty$ with $Q$ finite.  In this case, we consider again the magnitude 
(eq. \ref{bigspin8}).  $A$ can be written as $\lbrace 1+(\bar{\alpha}_{\mathrm{w}}^2-1)
\left[\frac{\lambda^2 Q^2}{N}- \frac{\lambda^4 Q^4}{3N^2}\right]\rbrace^{
\frac{N}{2}}$.  If $Q$ is finite when $N\rightarrow \infty$ then we can 
neglect  $\frac{1}{N^2}$ and higher terms from the expansion of $\sin^2$.  
Thus $A\approx \lbrace 1+(\bar{\alpha}_{\mathrm{w}}^2-1)\frac{\lambda^2 Q^2}{N}\rbrace^{
\frac{N}{2}}\approx\exp\lbrace \frac{(\bar{\alpha}_{\mathrm{w}}^2-1)\lambda^2 Q^2}{2}\rbrace$. 
 As long as $(\bar{\alpha}_{\mathrm{w}}^2-1)\lambda^2 < 1$, then the increase in  $A$, i.e.  $\exp \lbrace \frac{{\mathrm{i}}\lambda \hat{Q}_{\mathrm{md}}^{\mathrm{(N)}}}{\sqrt{N}} \sum_{\mathrm{j=1}}^N \tilde{\sigma}_{\mathrm{w}}^j  \rbrace$, is counter-balanced by the decline in the Gaussian $B$, and  thus eq. \ref{bigspin8} is centered around $Q=0$.

\subsubsection{\bf Finite N} 
\begin{figure}[here] 
{\includegraphics{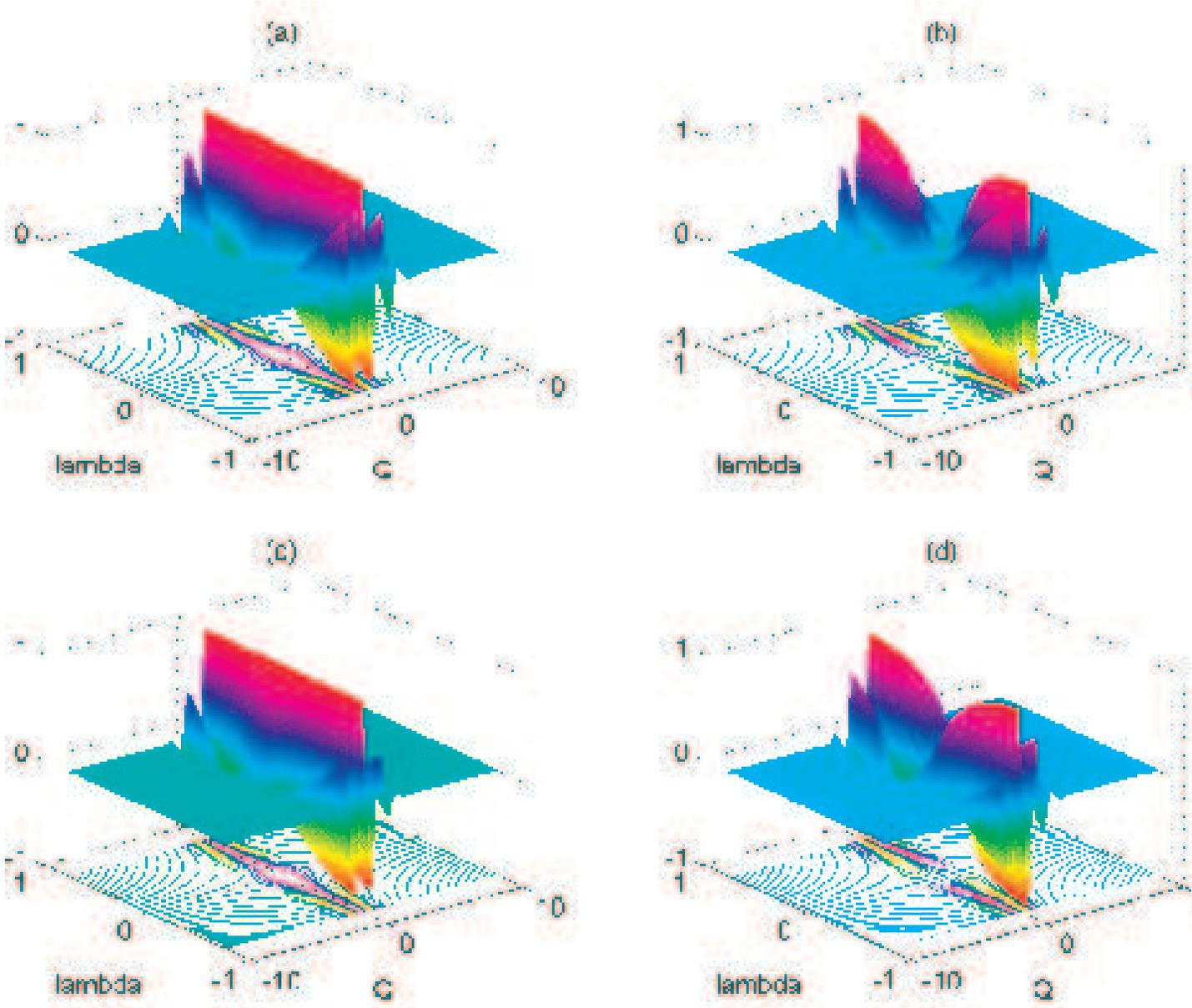}} 
\caption[Stern Gerlach Apparatus]{Numerical comparison of the RWM wavefunction (a-b) with the ideal shift by the WV $(\sigma_{45^{\circ}})_w=\sqrt{2}$ (c-d) for $N=20$. a) Real part of eq. \ref{bigspin6}, b) Imaginary part of eq. \ref{bigspin6}, c) Real part of eq. \ref{finalwm}, d) Imaginary part of eq. \ref{finalwm}.}\label{sgpre8} \end{figure}

We consider finite $N$ where there is a proper limit in which $N$ increases and the interaction goes to $0$ but $\lambda ^3$ is negligible.  If we fix $\lambda$ and choose an $N$ such that $\lambda \sqrt{N} > 1$, then we can measure the average exactly.  The uncertainty of $P_{\mathrm{md}}$ for $N$ particles is $\sqrt{N}$ and the momentum grows as $\lambda N \bar{\alpha} > \sqrt{N}$ which implies that $2\sqrt{N} \bar{\alpha} > 1$.  Nevertheless $N\lambda^3$ is still small (i.e. $N\lambda> \sqrt{N}$) but $N\lambda^3 < \frac{1}{\sqrt{N}}$ and $N\lambda^3$ is the extra correction.
For each spin there is a correction proportional to $\lambda ^3$ which for $N$ particles is $N\lambda^3$ which is small compared to $\sqrt{N}$ so $\sqrt{N} \lambda^3 < 0$ can be neglected.
We plot $N=20$ (fig. \ref{sgpre8}) to show that eq. \ref{finalwm}
 is an accurate approximation to eq. \ref{bigspin6}.

\subsubsection{\bf  MD shifted by WV}

Now $\sum_{\mathrm{j=1}}^N \bar{\alpha}_{\mathrm{w}}\approx N \bar{\alpha}_{\mathrm{w}}$ and thus 
all $N$ particles will deliver a momentum shift to the photographic plate of $\lambda N\bar{\alpha}_{\mathrm{w}}$ (and a shift in $\hat{P}_{\mathrm{md}}^{\mathrm{(N)}}$ of  $\lambda \sqrt{N}\bar{\alpha}_{\mathrm{w}}$ ).  The shift goes up as $N\bar{\alpha}_{\mathrm{w}}$  while the uncertainty goes up as ${\sqrt{N}}$ (the variance is 
$\langle (\Delta \hat{P}_{\mathrm{md}}^{\mathrm{(N)}})^2\rangle=[(\Delta \hat{P}_{\mathrm{md}})^2+\langle(\Delta (\sigma_\xi)_{\mathrm{w}})^2\rangle]$) and thus $\frac{\langle \hat{P}_{\mathrm{md}}^{\mathrm{(N)}}\rangle}{\langle \Delta \hat{P}_{\mathrm{md}}^{\mathrm{(N)}}\rangle}\approx\sqrt{N}$.   By choosing a sufficiently large $N$, the single trial WM outcome can be arbitrarily amplified. 
We have thus shown that it is valid to perform a WV approximation (i.e. to replace eq. \ref{bigspin6} with eq. \ref{finalwm}) in a significantly stronger coupling regime, i.e. for $\lambda \sim 1$,
by measuring a variable where it's shift is large compared to it's noise and thus, this is a RWM.  
\label{finiteq}


\subsubsection{\bf  Quantum Average of WVs} 

In the last 2 regimes, 
we chose, for simplicity, to ignore the details of a significant variation in the WVs,  
e.g. $\sum_{\mathrm{j=1}}^N \tilde{\sigma}_{\mathrm{w}}^j\approx N\bar{\alpha}_{\mathrm{w}}$.  
Even if the variance were significant, it is easy to see that our result is still valid.  I.e. even if the individual $\Psi$'s in the composite state are very different, then eq. \ref{avgop} is still valid, only the average will be over different pre- and post-selections.
 While it is appropriate to replace $\sigma$ by it's WV, 
there are 2 reasons that rotations can be induced in the pre- and/or post-selection.
Up to this point we have only discussed the first rotation (e.g. from $|{\uparrow_x} \rangle_j$  to $|\Psi \rangle_j$) which 
was corrected by the measurement of the relative positions.  This produced a shift in the pointer by an average over well-known WVs.
  However, $\exp \lbrace \frac{i\lambda \hat{Q}_{\mathrm{md}}^{\mathrm{(N)}}}{\sqrt{N}} \sum_{\mathrm{j=1}}^N \tilde{\sigma}_{\mathrm{w}}^j  \rbrace$ can cause a second rotation in the pre- or post-selection if  $\lambda\Delta Q_{\mathrm{md}}$ given by eq. \ref{bigspin7} is big enough.
However, in the instant case, the shift generated in the pointer 
$\exp \lbrace {\mathrm{i}}\lambda \hat{Q}_{\mathrm{md}}^{\mathrm{(N)}}\sqrt{N} \tilde{\sigma}_{\mathrm{w}}  \rbrace$ 
can be large, so even a small rotation to the pre- or post-selection will make a significant difference in the pointer shift.
We leave this analysis of  {\it additional} rotation in the pre- or post-selections  which cannot be determined by measurement of the relative positions to a future article.  However, we show here how 
 the total momentum, $\hat{P}_{\mathrm{md}}^{\mathrm{(N)}}$, is shifted by a quantum average of WVs~\cite{ab} due to the extra rotations by the additional uncertainty in $Q$ ( after the relative-position corrections are made)
with weights determined by the probability to obtain a particular $Q$ that is associated with a particular WV as suggested by eq. \ref{expweak}.  
As a simple example, one may  categorize the different WVs into different pre- or post-selections.  Suppose a subset, $n_1$, out of the ensemble of $N$ particles will all be rotated to the same state (e.g. to $|\Psi_1\rangle$) and thus will give one WV $\tilde{\eta}_{\mathrm{w}}^1$, other subsets will be rotated to another state (e.g. to $|\Psi_2\rangle$) giving another WV $\tilde{\eta}_{\mathrm{w}}^2$, etc.
Using $\sum n_i=N$.\footnote{For any product state, we still have that eq. \ref{thm1} is exactly true but with an $\bar{\sigma}=\sum {\tilde{\eta}}_n$, i.e. $\hat{Q}_{\mathrm{md}}^{\mathrm{(N)}}\hat{\sigma}\upn\ket{\Psi\upn}  = \hat{Q}_{\mathrm{md}}^{\mathrm{(N)}}\sum \tilde{\eta}_n\ket{\Psi\upn} + \hat{Q}_{\mathrm{md}}^{\mathrm{(N)}}\frac{\Delta
\sigma}{\sqrt{N}} |\Psi\upn\perp \rangle$.}
eq. \ref{bigspin6} is re-written:
\beq
\left[\exp \{ {\mathrm{i}}\lambda \hat{Q}_{\mathrm{md}}^{\mathrm{(N)}}\frac{n_1}{\sqrt{N}}\tilde{\eta}_{\mathrm{w}}^1\} 
\cdot\cdot\cdot\exp \{ {\mathrm{i}}\lambda \hat{Q}_{\mathrm{md}}^{\mathrm{(N)}}\frac{n_k}{\sqrt{N}}\tilde{\eta}_{\mathrm{w}}^n\}\right]\exp\left\{-{{({Q}_{\mathrm{md}}^{\mathrm{(N)}})^2}\over{4(\Delta {Q}_{\mathrm{md}}^{\mathrm{(N)}})^2}}\right\}
\label{toprove}
\eeq

\subsubsection{\bf  Obtaining EWVs instead of just an ordinary WV}.
In order to implement the amplification scheme, we need to obtain EWVs.  To obtain an EWV (i.e. outside the eigenvalue spectrum), rather than an ordinary WV, we need to control the rotation of the pre- or post-selection
by controlling $\lambda \Delta \hat{Q}$.
E.g. if $Q$ is limited (e.g. $\Delta Q_{\mathrm{md}}\approx 1$) and $\lambda$ is limited to a particular range sufficient to deliver a EWV  at every point of $Q$ within $\Delta Q_{\mathrm{md}}$, then we will also obviously obtain a EWV for the quantum average of WVs and do not need to be concerned with other issues such as the slope of $Q$.
However, anytime there is a way to get inside the spectrum of eigenvalues, there will be an exponential increase in the probability to obtain that WV.  
This can be seen from eq. \ref{expweak} in that  the fluctuation in the system is also relevant for the probability to obtain different  post-selections:
as the fluctuation in the system 
increases, the probability of a rare or eccentric post-selection also increases.  However an attempt  to see this through WMs will require 
the spread in the MD to be increased  and this increases the probability of seeing
the strange result as an error of the MD.  
\section{Conclusion}

In this article, we have introduced a new WM procedure for finite samples which yields accurate WVs that are outside the range of eigenvalues and which does not require an exponentially rare ensemble.
This procedure was motivated by an  application to
quantum metrology 
which provides a unique 
 advantage over the usual (pre-selected-only) approach in the amplification of small non-random signals: if the coupling $\lambda$ between system and MD is unknown and contains additional small errors that are not random, then actually performing a WM which yields an EWV can provide new information, e.g. by allowing us to distinguish  the shift from the non-random force (incorporated into $\lambda$) from the large EWV shift due to the WM interaction.  
The usual WV approximation becomes more and more precise in the idealized weak limit of $\lambda\Delta Q_{\mathrm{md}} \rightarrow 0$ and $N\rightarrow\infty$ in which there is no disturbance or back-reaction on the system.  Neither of these limits are realistic in practical amplifications because first of all, we must have a finite $N$, and second of all with a finite $N$, we must increase $\lambda$ to discern the WV from the noise.  RWM minimizes the two uncertainties in determining the WV which arise due to 1) the inability of MD to measure definite WVs due to the MD's uncertainty $\Delta P_{\mathrm{md}}$ and 2)  the back-reaction on the system due to $\Delta Q_{\mathrm{md}}$ creates an uncertainty in the WV of the system itself
By providing additional corrections for these uncertainties, RWM extends the coupling constant regime which thereby extends the potential utility of amplification of unknown forces~\cite{duck,abt}.

The RWM can also be used to augment the SWM of \S \ref{wvstat} given a large ensemble ($N\rightarrow\infty$) because we can now interpret what is the average of WVs corresponding to the total momentum shift for a stronger coupling constant by calculating the distribution of pre- and post-selections through the distribution of relative $Q$'s.  We can {\it calculate} what the distribution in $Q$ will be for $N\rightarrow\infty$, even if we do not know the distribution for the individual $Q$'s for any individual particles (the distribution of $\sum Q$ becomes a Gaussian for large $N$ for almost any individual distribution of $Q$). 
However, \cite{abt} for finite $N$ we cannot simply use a calculation because the fluctuation of the relative $Q$'s becomes important and can only be obtained through measurement.

Normally a valid WV calculation requires MD to be centered around $\hat{Q}_{\mathrm{md}}^{\mathrm{(N)}}=0$.  However, in \cite{ab} ideal measurements were converted to WMs  by post-selecting MD to be in a certain region of $Q$ and in $P_{\mathrm{md}}$, i.e. different regions of $\hat{Q}$ were sampled by multiplying by a function centered at $Q_{\mathrm{com}}$: i.e. a function of $Q'=Q-Q_{\mathrm{com}}$ such as  $\exp \frac{-(Q-Q_{\mathrm{com}})^2}{\Delta Q_{\mathrm{md}}^2}$ which is like starting the MD not with $Q=0$ but with $Q=Q_{\mathrm{com}}$.  Results centered at different $Q=Q_{\mathrm{com}}$ are  then summed.   However, even such limited projections can still disturb each other.   The new RWM presented here is more subtle because the relative coordinates commute with the total momentum and so can be simultaneously measured without disturbing each other.  
By measuring the relative positions, we can go  beyond the weak approximations  used in the past (i.e. $\lambda\ll 1$).
Since we are able to measure the relative positions exactly, we are also able to make these corrections exactly.  
We thus have a much stronger interaction (i.e. a $\lambda $ that does not have to be $\ll 1$) and still we can 
obtain EWVs.

{\bf Acknowledgments:} The author thanks Yakir Aharonov and Alonso Botero for many fascinating discussions.


\begin{thebibliography}{99}
\bibitem{avg}

{Theorem:} for every observable $A$ and a normalized state $|\psi\rangle$, we have:
 $A|\psi\rangle=\langle  A\rangle|\psi\rangle+\Delta A|\psi_\perp\rangle$
   for some state $|\psi_\perp\rangle$ which is orthogonal to $|\psi\rangle$.
To prove this, we begin with:
$A|\psi\rangle=\langle A\rangle|\psi\rangle+A|\psi\rangle-\langle A\rangle|\psi\rangle$
   now, we set:
$|\widetilde{\psi}_\perp\rangle =A|\psi\rangle-\langle A\rangle|\psi\rangle$,   so: $\langle \widetilde{\psi}_\perp |\psi\rangle =(\langle\psi|A-\langle\psi|\langle
      A\rangle)|\psi\rangle=\langle\psi| A|\psi\rangle - \langle A\rangle\langle\psi|\psi\rangle = 0$
   now we set: $|\psi_\perp\rangle=b|\widetilde{\psi}_\perp\rangle$, where $|\psi_\perp\rangle$ is normalized and $b$
   real (note that $\langle\psi|\psi_\perp\rangle=0$).
   so:
$A|\psi\rangle=\langle A\rangle|\psi\rangle+b|\psi_\perp\rangle$.
   Now we multiply from the left by $\langle\psi_\perp|$, and we get:
   $\langle\psi_\perp|A|\psi\rangle=b$.   Now we can see that:
$\langle\psi|A^2|\psi\rangle=\langle\psi|A(\langle A\rangle|\psi\rangle+b|\psi_\perp\rangle)=
      \langle\psi|({\langle A\rangle}^2|\psi\rangle+b\langle A\rangle|\psi_\perp\rangle+bA|\psi_\perp\rangle)=
      {\langle A\rangle}^2+b\langle\psi|A|\psi_\perp\rangle$  so: $\langle A^2\rangle-{\langle A\rangle}^2=b\langle\psi|A|\psi_\perp\rangle=b^2$
   which means that: $b=\sqrt{\langle A^2\rangle-{\langle A\rangle}^2}=\Delta A$
   and the result: 
   $A|\psi\rangle=\langle  A\rangle|\psi\rangle+\Delta
   A|\psi_\perp\rangle$  is proved.
\bibitem{proof} The identity $\exp\left\{{i\alpha\hat{\sigma}_{\vec{n}}}\right\}=\cos\alpha+i\hat{\sigma}_{\vec{n}}\sin\alpha$ is easily proven using the fact that for any integer $k$:
$\sigma_{\hat{n}}^{2k}=I$ and $\sigma_{\hat{n}}^{2k+1}=\sigma_{\hat{n}}$
   and now it follows that:
$      e^{i\alpha\sigma_{\hat{n}}}=\sum_{k=0}^\infty\frac{(i\alpha)^k\sigma_{\hat{n}}^k}{k!}=\sum_{k=0}^\infty\frac{(i\alpha)^{2k}}{(2k)!}+\sigma_{\hat{n}}\sum_{k=0}^\infty\frac{(i\alpha)^{2k+1}}{(2k+1)!}
      =e^{i\alpha\sigma_{\hat{n}}}=\cos\alpha+i\sigma_{\hat{n}}\sin\alpha$  and the identity is proven.
\end{thebibliography}
\end{document}